\documentstyle[12pt,psfig]{article}
\begin{document}

\title{MICROLENSING}
\author{Esteban Roulet\thanks{Email: roulet@susy.sissa.it} and
Silvia Mollerach\thanks{Email: mollerac@neumann.sissa.it} \\
{\it International School for Advanced Studies, SISSA}\\
Via Beirut 2, I-34013, Trieste, Italy}
\date{}
\maketitle
\vfill

\begin{abstract}
Microlensing observations have now become a useful tool in searching
for non--luminous astrophysical compact objects (brown dwarfs, faint
stars, neutron stars, black holes and even planets). Originally
conceived for establishing whether the halo of the Galaxy is composed
of this type of objects, the ongoing searches are actually also
sensitive to the dark constituents of other Galactic components (thin
and thick disks, outer spheroid, bulge). We discuss here the present
searches for microlensing of stars in the Magellanic Clouds and in the
Galactic bulge (EROS, MACHO, OGLE and DUO collaborations). We analyse
the information which can be obtained regarding the spatial
distribution and motion of the lensing objects as well as about their
mass function and their overall contribution to the mass of the
Galaxy. We also discuss the additional signals, such as the parallax
due to the motion of the Earth, the effects due to the finite source
size and the lensing events involving binary objects, which can
further constrain the lens properties. 
 We describe the future prospects for these searches and the further
proposed observations which could help to elucidate these issues, such
as microlensing of stars in the Andromeda galaxy, satellite parallax
measurements and infrared observations. 
\end{abstract}
\vfill
\leftline{To appear in Physics Reports}
\leftline{SISSA--171/95/EP}
\pagebreak
\
\tableofcontents
\pagebreak

\section{Introduction}

The bending of light in a gravitational field predicted by general 
relativity provided one of the first verifications of Einstein 
theory \cite{ed19}. 
The value of the deflection angle which results has now been tested 
at the 1\% level by observing 
the change in the apparent position of stars whose light is deflected
by the Sun. In 1920, Eddington \cite{ed20} noted that the light deflection 
by a stellar object would lead to a secondary dimmer image 
of a source star on the opposite side of the deflector. Chwolson 
\cite{ch24} later pointed out that these secondary images could make
foreground stars appear as binaries, and that if the alignment were 
perfect, the image of the 
source would be a ring. In 1936, 
Einstein \cite{ei36} published the correct formula for the 
magnification of the two images of a very distant star, and concluded 
that this lensing effect was of no practical relevance due to the 
unresolvably small angular separation of the images and the low 
probability for a high amplification event to take place. 

The following year, Zwicky \cite{zw37} showed that, if the deflecting 
object were a galaxy instead of a single star, the gravitational 
lensing  of the light of a background galaxy would lead to 
resolvable images. This `macrolensing' effect would 
provide information about the mass of the intervening galaxy and allow  
observation of objects at much larger distances due to the magnification of 
their light. Furthermore, the probability that this effect be observed 
is a `certainty' \cite{zw37b}.

It was actually the multiple imaging of a high redshift quasar by a
foreground galaxy which provided the first observation of
gravitational lensing 
\cite{wa79}, and this area of research has now become an active field in 
astronomy, with the potential of giving crucial information for 
cosmology \cite{sc92,bl92,re94}. For instance, the time delay among the 
multiple images of a quasar allows one to relate the lens mass to the 
Hubble constant. Another interesting example is that the mass 
distribution of a rich cluster can be reconstructed from the shapes of 
the images of 
thousands of background faint blue galaxies around it, which become 
elongated due to the effect of weak lensing by the cluster \cite{ty95}. 
Also, gravitational lensing
 can strongly affect the observable properties of AGN 
and quasars, and may have to be taken into account when inferring their 
intrinsic properties.

While these developments were going on, the theoretical study of
gravitational lensing  of 
stars by stars restarted in 1964, with the work of Liebes \cite{li64} 
and Refsdal \cite{re64}, who extended the formalism and discussed the 
lensing of stars in the disk of the Galaxy, in globular clusters and in the 
Andromeda galaxy.

The lensing effect of individual stars belonging to a galaxy that is 
itself macrolensing a background source was discussed by Chang and 
Refsdal \cite{ch79}. When both the lensed source and the intervening 
galaxy are at cosmological distances, the passage of one of these 
stars close to the line of sight to one of the images  further 
deflects the source light 
 by an angle which is typically of ${\cal O}(\mu$arcsec). The name 
`microlensing' (ML) then became 
 associated with this process, and is now generally 
applied to any gravitational lensing effect by a 
compact object producing unresolvable images of a source and 
potentially huge magnifications of its light.

Press and Gunn \cite{pr73} showed in 1973 that a cosmological density 
of massive compact dark objects could  manifest through the ML of high 
redshift sources.
In 1981, Gott \cite{go81} pointed out the possibility of detecting the 
dark halos of remote galaxies by looking for ML of background 
quasars. It was in 1986 that Paczy\'nski \cite{pa86} gave a new face 
to the field when he noted that, by monitoring the light--curves of 
millions of stars in the Large Magellanic Cloud (LMC) for more than a 
year, it would become possible to test whether the halo of our galaxy 
consisted of compact objects with masses between $10^{-6}$ and 
$10^2\ M_\odot$, i.e. covering 
most of the range where baryonic dark matter in the form of planets, 
Jupiters, brown dwarfs or stellar remnants (dead stars, neutron stars 
or black holes) could lie.
It was later realized \cite{pa91,gr91b}
that the Galactic bulge stars also provided an 
interesting target to look at, since at least the lensing by faint 
stars in the disk should grant the observation of ML events, and 
observations in the bulge  
could also allow one to test the dark constituency of the Galaxy close to 
the Galactic plane.

With these motivations, several groups undertook the observations 
towards the LMC (EROS and MACHO collaborations) and towards the bulge 
(MACHO and OGLE), obtaining the first harvest of ML events in 1993
\cite{al93,au93,ud93}. As will be discussed in this review, 
the observations at present are already providing crucial insights 
into the dark matter problem, into the non--luminous contribution to 
the mass of the different Galactic components 
 and they are helping to unravel the morphology of 
the inner part of the Galaxy. The continuation of these programs, as 
well as the new ones entering the scene (DUO collaboration looking
towards the bulge; AGAPE and the Columbia--VATT collaborations 
looking at Andromeda and followup telescope 
networks such as PLANET, GMAN and MOA
looking for fine details in the microlensing events) will certainly allow 
us to reach a much deeper knowledge of these fundamental issues in the 
near future.

\section{Galactic components and  dark matter}

In order to study the lensing effects of objects in and around the 
Galaxy, it is first convenient to discuss the nature of the several 
possible lens candidates as well as their distribution and expected 
properties.
In this Section we briefly introduce the different Galactic components
\cite{fi91,fr87,ba86b} and the possible lensing populations present in
them.

\subsection{Dark halos}
The existence of dark galactic halos is directly related 
 to the dark matter (DM) issue \cite{as92}, namely the 
fact that more than 90\%, and probably even 99\%, of the matter in the 
Universe does not seem to emit any electromagnetic radiation. Its 
existence has been inferred from the gravitational effects that it has on 
the observed motions of gas, stars, galaxies and clusters, and also 
from its effects on the overall geometry of space--time.  

For instance, 
the velocity dispersions of galaxies in clusters indicate that 
cluster masses are more than 10 times larger than the mass contained 
in the luminous parts of its constituent galaxies and in the
intra--cluster gas \cite{zw33,wh93}. Similar 
conclusions are obtained from the X--ray observations of the cluster gas 
temperature profiles \cite{wh93} as well as from weak gravitational 
lensing \cite{ty95}. At Galactic scales, the presence of DM is most
clearly indicated by the flatness of the rotation curves of spiral galaxies
\cite{sa87,pe95},  and
also from the motion of satellite galaxies \cite{za94}. 

The existence of dark halos surrounding the 
visible galaxies is supposed to account, at least in part, for 
this missing mass (see e.g. Ref. 
\cite{os74}). The halo densities are 
 then required to fall as $r^{-2}$ with the galactocentric distance
$r$,  as long as the rotation curves stay flat. A  
parameterization usually adopted for the density of 
a spherically symmetric halo is 
\begin{equation}
\rho^H(r)=\rho^H_0{(R_0^2+a^2)\over(r^2+a^2)},
\label{rhoh.eq}
\end{equation}
where $a$ is the halo core radius, typically a few kpc. For our
galaxy, taking  $R_0$ as the solar galactocentric distance, we have that in
Eq.~(\ref{rhoh.eq})  $\rho_0^H$ is the 
local halo density. In the following we will adopt $R_0=
8.5$~kpc and we will take the reference value
$\rho_0^H\simeq 10^{-2}M_\odot/$pc$^3$ \cite{ca81,ba86b,ga95b}.
 
If we assume  that the Milky Way circular speed $v_c\simeq 220$~km/s
 stays constant up to a galactocentric 
distance $r_{max}$, the total  mass of the Galaxy 
within that radius results ($G$ is Newton's constant)
\begin{equation}
M_{tot}\simeq {v_c^2r_{max}\over G}\simeq 5.6\times 10^{11}
\left({v_c\over 220\ {\rm 
km/s}}\right)^2\left({r_{max}\over 50\ {\rm kpc}}\right)M_\odot,
\label{mhalo.eq}
\end{equation}
which is an order of magnitude larger than the disk mass 
if $r_{max}>50$~kpc, and hence should mainly be due to the halo.
The large amounts of matter contained in the dark Galactic halos may 
account for the missing mass in clusters as well. 

The detailed profile and extent of galactic halos, in particular that of 
the Milky Way, are however not well known, partly due to 
the poor knowledge of the rotation curves. For our galaxy, the
rotation curve has 
been measured only up to $\sim 20$ kpc and with large uncertainties 
\cite{fi91}.
 There is also an 
uncertainty affecting the halo density in the inner region of the 
Galaxy which is related to the imprecise knowledge of the contribution 
to the rotation curve coming from the Galactic disk. For instance, 
maximum disk models \cite{se88} would lead to little halo mass inside the 
solar galactocentric radius.

Another poorly constrained property of halos is their ellipticity
\cite{ri95}  
which, for instance, in simulations of galaxy formation with cold dark 
matter turns out to be non--negligible. Observations of 
polar ring galaxies, in which the motion of gas in a ring orthogonal
to the plane of the disk can 
constrain  the 
flattening of the mass distribution, have also provided some examples of 
flattened halos. However, the 
flaring of HI in the outskirts of the disks in spiral 
galaxies indicates that the halo is not distributed as a
thin disk \cite{ku82,me92}.
A further complication is that a flattened 
halo may not be axisymmetric and may 
also be tilted with respect to the disk plane.

Associated with all these uncertainties there is also an uncertainty in 
the motion of the halo constituents. The simplest hypothesis that is 
usually adopted is that the halo objects move with isotropic 
velocities, which have a Maxwellian distribution with constant (space 
independent) one--dimensional velocity dispersion $\sigma$, which is 
then related to the circular speed by $\sigma=v_c/\sqrt{2}\simeq
155$~km/s.  This 
so--called isothermal sphere is known to lead to a density profile  
falling as $r^{-2}$ at large radii.  A self--consistent halo 
distribution, not including the effects of the disk but allowing for 
an elliptical halo as well as for a rising or declining rotation curve, 
has been obtained in Ref. \cite{ev94}. Some progress towards 
including the dynamical effects of disks and bulges has also 
recently been made \cite{ku95}.

\subsection{Baryonic or non-baryonic?}

The nature of the dark halo constituents is one of the important open 
problems of present--day physics. One possibility is that they are
composed  of 
just ordinary baryonic matter, e.g. protons and neutrons. 
A very important bound for the total amount of baryons in the Universe 
comes from the agreement between the observed abundances of light 
elements (D, $^3$He, $^4$He, $^7$Li) and those predicted in the 
context of the primordial nucleosynthesis theory.  This requires that 
the average density of baryons $\rho_b$, in terms of the critical density of 
the Universe $\rho_c$\footnote{$\rho_c\equiv 3H_0^2/(8\pi G)$ is the
density leading to a flat Universe, with $H_0$ being the Hubble
constant.}, 
$\Omega_b\equiv\rho_b/\rho_c$, 
 be in the range $0.009<\Omega_b<0.14$ \cite{co95}. 
Since luminous matter accounts 
for $0.003<\Omega_{lum}\le 0.007$ \cite{pa90,pe92}, 
the lower bound on $\Omega_b$ suggests 
the existence of baryonic matter in dark forms. The upper bound on 
$\Omega_b$ is consistent with dark baryonic halos accounting for the
observations at galactic scales, which require $\Omega\geq
0.1$. However,  when 
 combined with observations on cluster scales or larger 
\cite{be89}, which imply that $\Omega_{tot}>0.2$--0.3 (and probably 
$\Omega_{tot}\simeq 1$, which is the value preferred by  inflationary 
theories), the above mentioned upper bound, $\Omega_b<0.14$, suggests 
the existence of non--baryonic dark matter 
(NBDM). 

An additional attractive feature of NBDM is that it proves 
very useful in allowing for the formation of structure, in scenarios where 
the inhomogeneities in the density grow first in the NBDM, with the 
baryons falling later into the dark potential wells. In these scenarios, 
the NBDM that seeded galaxy formation is expected to remain 
somewhere in the galaxies, with the halo being the most `natural' 
place. Candidates for NBDM abound in the literature, going from 
primordial black holes (made essentially out of radiation) to elementary 
particles such as massive neutrinos, axions, supersymmetric 
neutralinos, etc..

On the other hand, as mentioned above, halos may well be baryonic and there
are actually 
 many ways in which baryons can hide in dark forms, without entering
into 
conflict with  other observations (for a review see Ref. \cite{ca94}). 
These are:
\begin{itemize}
\item Very massive black holes (with mass 
$m>200\ M_\odot$, in order to avoid the 
ejection of too many heavy elements into the interstellar medium).
\item
Stellar remnants, such as neutron stars or dead 
white dwarfs ($m\sim 
M_\odot$).
\item Brown dwarfs, which are stars made essentially 
of H and He that 
are too light to start efficient nuclear fusion reactions 
($m<m_{HB}\simeq 0.08\ M_\odot$, where $m_{HB}$ is the minimum star mass 
for H burning to take place). The Jeans mass, i.e. the minimum mass 
for which the gravitational forces of an isolated gas cloud fragment 
can overcome the pressure forces and lead to star formation, has been 
estimated to be 4--$7\times 10^{-3}\ M_\odot$ \cite{lo76,pa83}. 
Under special conditions (e.g. in protoplanetary disks), this lower 
bound can be evaded. Hydrogenous objects with masses in the neighbourhood of 
$10^{-3}\ M_\odot$ are usually named Jupiters. 
\item Snowballs, which are 
compact objects with masses below $10^{-3}\ M_\odot$, held 
together by molecular, rather than gravitational, forces.
A lower bound of $10^{-7}\ M_\odot$  for their masses 
has been suggested in order to avoid their evaporation \cite{de92}.
\item Clouds of molecular H have been proposed as baryonic 
constituents of the dark Galactic halos \cite{pf94,de95,ge95}.
\end{itemize}
These possibilities, with the exception of the last one, 
clearly provide 
excellent candidates for gravitational lenses. These objects
are sometimes generically called Massive Compact Halo Objects
(MACHOs). Of course nothing is known about the mass function of the 
halo constituents, except that the number of ordinary main sequence stars
(see Section~6.3) and  of
white dwarf remnants \cite{ch95} should be quite suppressed.

\subsection{Galactic disk}
Most of the visible stars in our galaxy are distributed in a disk, 
generally assumed to be a double exponential in the galactocentric 
cylindrical coordinates ($R,{\rm z}$)
\begin{equation}
\rho^D(R,{\rm z})=\rho^D_0\;{\rm exp}\left(-{R-R_0\over h_R}-{|{\rm z}|\over 
h_{\rm z}}\right),\label{rhod.eq}
\end{equation}
where $\rho_0^D$ is the local disk density, $h_R\simeq 3.5$~kpc is the 
disk scale length \cite{ba86b} (estimates of $h_R$ vary however in the wide
range from 1.8 to 6~kpc \cite{ke91}) and $h_{\rm z}$ is the scale
height, which is $\sim$100~pc for the 
very young stars and gas, and $\sim$325~pc for the older disk 
stars \cite{ba86}. 
There are some indications \cite{pa94b} 
that at distances larger than $\sim 3$~kpc 
towards the Galactic center the disk 
star density falls below what would follow from Eq.~(\ref{rhod.eq}),
suggesting that the scale height may decrease with decreasing $R$
\cite{ke91} or that the disk is actually  `hollow' 
($\rho^D\simeq 0$ for $R<2$--4~kpc). Also Eq.~(\ref{rhod.eq}) does not
describe the existence of a spiral structure, which is probably
responsible for an excess in the star counts at distances $\sim 2$~kpc
from the Sun (Sagitarius arm) in the direction of the Galactic center.

The disk is supported by the rotation of its stars, 
$v^D_{rot}\simeq 220$~km/s. 
The velocity dispersion $\sigma^D$ of disk stars is small: 
 $\sigma^D\simeq 20$~km/s locally, although it is
  probably larger at smaller $R$ \cite{le89}. The 
existence of a somewhat hotter component, the thick disk, was 
suggested by Gilmore and Reid \cite{gi83}. This
component has velocity dispersion 
$\sigma^{TD}\simeq 40$~km/s, slower rotation ($v_{rot}^{TD}\equiv 
v_{rot}^D-v^{TD}_{drift}$, with the asymmetric drift being 
$v^{TD}_{drift}\simeq 40$~km/s) and larger scale height, 
$h^{TD}_{\rm z}\simeq 1.2$~kpc \cite{fr87}. 
While the `thin' disk is made of relatively young metal 
rich stars, the thick disk stars have smaller metallicities, and so
are probably older. 
It is not yet clear whether the thick disk is just the 
old tail of the thin disk or is instead a separate component formed 
independently.

The thin disk local column density, $\Sigma_0^D\equiv 2h_{\rm
z}\rho_0^D$, has been
measured in several ways. Bahcall \cite{ba84} initially obtained 
for the observed matter $\Sigma_0^D\simeq 
48\ M_\odot/$pc$^2$, with $\sim 10\ M_\odot/$pc$^2$ in gas,
$\sim 5\ M_\odot/$pc$^2$ in white dwarfs and red giant stars and the
rest in main sequence stars. A more recent analysis \cite{go96} finds
a  smaller M dwarf contribution, leading to $\Sigma_0^D\simeq
40 M_\odot$/pc$^2$.
On the other hand, Kuijken and Gilmore
\cite{ku91} obtain, from the study of the vertical motion of stars,
that the total (disk+halo) column density within $\pm 1.1$~kpc of the
disk plane is
$\Sigma_0(|{\rm z}|<1.1$~kpc$)=(71\pm 6)M_\odot/$pc$^2$. Removing the halo
contribution with a rotation curve constraint they obtain, for the disk
alone,  $\Sigma^D_0(|{\rm z}|<1.1$~kpc$)=(48\pm 9)M_\odot/$pc$^2$. A
reanalysis by Gould of the same data lead to 
$\Sigma_0^D(|{\rm z}|<1.1$~kpc$)=(54\pm 8)M_\odot/$pc$^2$ \cite{go90} (see
also Ref.~\cite{ba92}). This does
not suggest the presence of large amounts of DM in the disk. This is
further supported by the fact that the local mass distribution of main
sequence stars \cite{sc86,ti93,go96} 
has been measured almost down to the hydrogen burning limit, 
and it is consistent with being approximately 
flat below $m\simeq 0.4 M_\odot$, so that 
a naive extrapolation of it below $m_{HB}$ does not hint towards a
significant contribution to the disk mass coming from the brown dwarfs.
If we take as an upper limit for the disk column density within
$\pm 1.1$~kpc the value 70$M_\odot$/pc$^2$, and subtract from it the column
density of known matter, 
this would allow a maximum
dark contribution to the thin disk density of $\Sigma_0^D\simeq 30\
M_\odot$/pc$^2$. For the thick disk, $\Sigma_0^{TD}\simeq 45\ M_\odot$/pc$^2$
may be allowed, since for a distribution with scale height $\sim
1$~kpc only 2/3 of the total column density would be within those
heights. 

Finally, let us note that the total mass associated with the disk, in
terms of the local column density $\Sigma_0$ and assuming the scale
length to be 3.5~kpc, is (upon integration of Eq.~(\ref{rhod.eq}))
\begin{equation}
M^D=4.4\times 10^{10}M_\odot\left({\Sigma_0\over 50\ M_\odot{\rm
pc^{-2}}}\right).
\end{equation} 

 The possible ML effects of disk populations for observations
towards the LMC were first discussed in  Ref.~\cite{go94,go94c}. For
observations in the bulge direction see Ref.~\cite{pa91,gr91b}. 

\subsection{Galactic spheroid}

The other population of stars, which is observed both looking at high
latitudes far above the disk and locally by studying the high velocity
stars which cannot be bound to the disk,
is the Galactic spheroid (sometimes referred to by astronomers as the
`stellar halo'). It is a population of old metal poor stars,
probably of protogalactic origin, with little rotation, supported by a
large velocity dispersion $\sigma^S\simeq 120$~km/s and density
falling as $r^{-3.5}$. This is consistent with the expectation
$\sigma=v_c/\sqrt{n}$ for an isotropic component with density falling
as $r^{-n}$ in a potential leading to a constant rotation velocity
$v_c$ \cite{ri90}. There may however be departures from sphericity 
in the spheroid density and from isotropy in the 
velocity distributions \cite{ba86,bi86,bl91a}.

The local density in luminous spheroid stars, i.e. those more massive
than $m_{HB}$,  was estimated by Bahcall,
Schmidt and Soneira \cite{ba83} to be $\rho^S_0(m>m_{HB})\simeq 10^{-4}
M_\odot$/pc$^3$. This implies a luminous spheroid which is
dynamically irrelevant. However, dynamical models of the Galaxy, in
which the same spheroid population  accounts for the dynamic and
photometric observations in
the inner Galaxy, suggested already many years ago that the spheroid
mass may be much larger, with $\rho^S_0\simeq 10^{-3}M_\odot$/pc$^3$
\cite{ca81,os82,ro88}. This difference between luminous and dynamical
estimates may be due to the presence of large amounts of stars with
masses below $m_{HB}$, as seems to be suggested by the measured mass
function of spheroid field stars, which rises as $dn/dm\propto
m^{-4.5\pm 1.2}$ below 0.5~$M_\odot$ and down to the lowest measured
value of 0.14~$M_\odot$ \cite{ri92}. Also, a sizeable number of stellar
remnants could contribute to these unseen matter.
These objects could  provide an
interesting population of lensing agents \cite{gi94}.

When discussing the heavy spheroid models, we will adopt a simplified
density profile which fits well the Ostriker and Caldwell model
\cite{os82}, given by
\begin{equation}
\rho^S(r)=\rho^S_0\left({\sqrt{R_0}+\sqrt{b}\over
\sqrt{r}+\sqrt{b}}\right)^7,
\label{rhos.eq}
\end{equation}
where $\rho^S_0=1.5\times 10^{-3}M_\odot/$pc$^3$ and the core radius is
taken as $b=0.17$~kpc. The total mass of this spheroid model is
$5.7\times 10^{10}M_\odot$.

\begin{figure}
\centerline{\psfig{figure=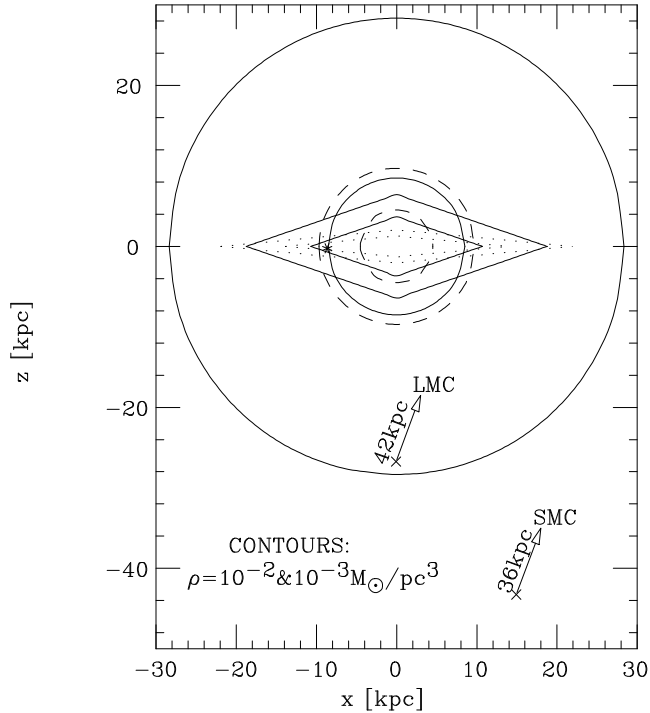,angle=90,height=10cm}}
\caption{Schematic view of the possible Galactic lensing populations:
a standard spherical halo (solid lines), a heavy spheroid (dashed
lines), a maximum thick disk (solid lines) and dark thin disk (dotted
lines). The density contours $\rho=10^{-3}$ and
$10^{-2}M_\odot$/pc$^3$ are shown. The locations of the Sun, the LMC
and the SMC are indicated.}
\label{map.fig}
\end{figure}

 Figure~\ref{map.fig} shows a schematic view of the above mentioned
Galactic components. 
The contours $\rho=10^{-2}$ and $10^{-3}  M_\odot$/pc$^3$
are plotted for the halo (assuming $\rho_0^H=10^{-2} M_\odot$/pc$^3$
and $a=3$~kpc), for allowed dark disk components with $\Sigma^D_0=30
 M_\odot$/pc$^2$ and $\Sigma^{TD}_0=45 M_\odot$/pc$^2$ (the luminous
contribution to the thick disk column density is much
 smaller than this), and for the heavy spheroid model of
Eq.~(\ref{rhos.eq}). The location of the Sun, 
at ($-8.5$,0,0)~kpc, is indicated with a star.
The location of the Large and Small Magellanic Clouds is also shown,
assuming that they are at 50~kpc and 60~kpc respectively, and using
their Galactic coordinates $(\ell,b)_{LMC}=(281^\circ,-33^\circ)$ and
$(\ell,b)_{SMC}=(303^\circ,-44^\circ)$.

\subsection{Galactic bulge}

The inner 1--2~kpc of the Galaxy are usually referred to as the
Galactic bulge. 
The bulge is sometimes
considered just as the inner part of the spheroid
mentioned above \cite{ca81,bl89}, 
and at some other times, instead, as a new unrelated
central component \cite{ba83}.
 There are actually indications that only
the metal poor stars in the bulge 
have spheroid--like kinematics, while there is  an
additional population of stars with larger metallicities \cite{mc94}
 which is more
centrally concentrated, has a velocity dispersion which decreases with
distance and has a non--negligible rotation \cite{ri90,mi95}.
The origin of this component, and  its relation
to the spheroid and disk,  
is still under debate \cite{ri92b,wy92,mi95}, but it was 
 probably formed from the
dissipated gas left out after the spheroid
formation, eventually affected by a bar instability of the inner disk.

Due to the high extinction in the direction of the Galactic center,
observations of the bulge are performed in some clear windows (for
example in Baade's Window  at $\ell=1^\circ$ and $b=-3.9^\circ$), 
or at infrared wavelengths. In recent years,
reanalyses of infrared observations of Matsumoto et al. by Blitz and
Spergel \cite{bl91b}, of Spacelab data by Kent \cite{ke92} and
COBE--$FIRAS$ 
data by Dwek et al. \cite{dw95}, have provided evidence that the bulge
is not spherically symmetric, probably having the shape of a bar
\cite{bl91b,dw95} with
its near side in the first Galactic quadrant. Besides the results from
the integrated IR light, similar results arise from star count
studies (for a review see \cite{sp92}), and from the study of gas
motions \cite{bi91}.

Two models are commonly used to describe the inner Galaxy. The first
is the axisymmetric bulge model of Kent \cite{ke92}, in which (scaling
his results to $R_0=8.5$~kpc)
\begin{equation}
\rho^K(t)=\left\{\begin{array}{ll} 0.75\left(t/{\rm
kpc}\right)^{-1.85}{M_\odot/{\rm pc}^3}&,t<1\ {\rm kpc}\\
3.13\,K_0\left({1.41 t/{\rm kpc}}\right){M_\odot/{\rm pc}^3}&
,t>1\ {\rm kpc},
\end{array}\right.
\label{kent}
\end{equation}
where $t^4=R^4+({\rm z}/0.61)^4$ and $K_0$ is a Bessel function.

The second is the model for the bar obtained by Dwek et
al. \cite{dw95}, which is given by
\begin{equation}
\rho^B(s)={M_0\over 8\pi abc}{\rm exp}\left[-{s^2\over 2}\right]
\label{dwek}
\end{equation}
where 
$s^4\equiv \left[({\rm x'}/a)^2+({\rm y'}/b)^2\right]^2+({\rm
z'}/c)^4$, with 
the primed axes being along the principal axes of the bar. These
coordinates  are
related to the galactocentric coordinates in which x points opposite
to the Sun's direction, y towards the direction of increasing
longitudes and z to the north Galactic pole, by
${\rm x'=x\cos\alpha+y\sin\alpha}$ and ${\rm
y'=-x\sin\alpha+y\cos\alpha}$, 
with $\alpha\simeq 20^\circ$ being 
the angle between the major axis of the bar
and the x axis, and ${\rm z'=z}$. The scale lengths of the bar are
$a=1.58$~kpc, $b=0.62$~kpc and $c=0.43$~kpc.

\begin{figure}
\centerline{\psfig{figure=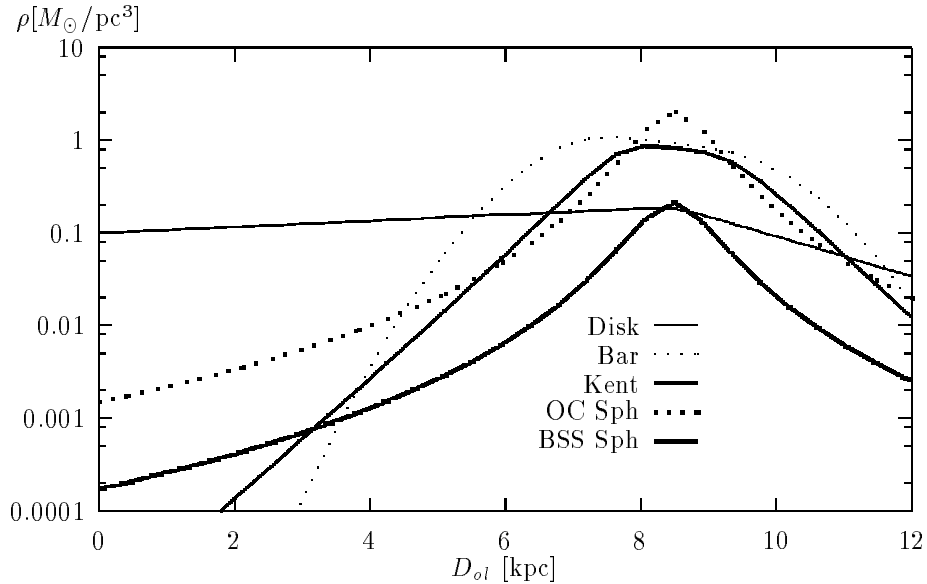,height=10cm}}
\caption{Density profiles in the direction of Baade's Window for the
different lensing populations considered, as specified in the text.}
\label{rho.fig}
\end{figure}

Figure \ref{rho.fig} shows the density profiles towards Baade's Window
 for these models. We take a bar mass $M_{bar}\equiv 0.82 M_0=2\times
10^{10}M_\odot$, in the upper range of the dynamical estimates
$M_{bar}=1$--$2\times 10^{10}M_\odot$ (see however Ref. \cite{bl95}).
 We also show the thin disk density profile, assuming
$\rho_0^D=0.1 M_\odot$/pc$^3$ (corresponding to 
$\Sigma_0^D\simeq 65M_\odot/$pc$^2$); 
 the heavy spheroid density of the 
Ostriker and Caldwell model~\cite{os82}; as well as the `light'
spheroid density of the
Bahcall, Schmidt and Soneira model~\cite{ba83}.

We note that the Kent and bar models are not intended to describe the
star distribution beyond $r\sim 2$~kpc, where they become exponentially
suppressed\footnote{Dwek's model is also not appropriate for describing the
inner 500~pc of the bulge \cite{ha95b}.}. In this sense, they play the 
role of the central component
invoked by Bahcall, Schmidt and Soneira 
to account for the inner Galactic observations. On the
other hand, the inner part of the heavy spheroid model provides a
crude approximation to Kent's model in the central 2~kpc, but they
differ considerably in the outer parts.  Finally, we note that
considering spheroid models with larger core radius ($b\geq 1$~kpc), as
done in \cite{ga95c}, one may
still have a heavy outer spheroid, which should then be essentially
dark, but which contributes only a little to the mass of the bulge.

Regarding the velocity distribution of bulge objects, the average
line of sight velocity dispersion of stars in Baade's Window
 is $\sigma\simeq 110$~km/s
\cite{ri90}. For the bar model of Dwek et al., which should have an
anisotropic velocity ellipsoid, Han and Gould \cite{ha95} obtained, using the
tensor virial theorem and normalizing to the observed
line of sight dispersion,  that the velocity dispersions along the bar
axes are ($113.6,\ 77.4,\ 66.3$)~km/s for an assumed inclination
$\alpha=20^\circ$. Detailed modeling of the star orbits in the bar,
and a discussion of the implications for ML observations, has
been made by Zhao, Spergel and Rich \cite{zh95,zh96}.

\section{Microlensing formalism}
We discuss in this Section the main features of the ML events, and
present the formalism required for describing the
ML  observations \cite{ei36,li64,re64,pa86,ne91,gr91,de91,de95b,ha95}.
 
\subsection{Light--curves}

\begin{figure}
\centerline{\psfig{figure=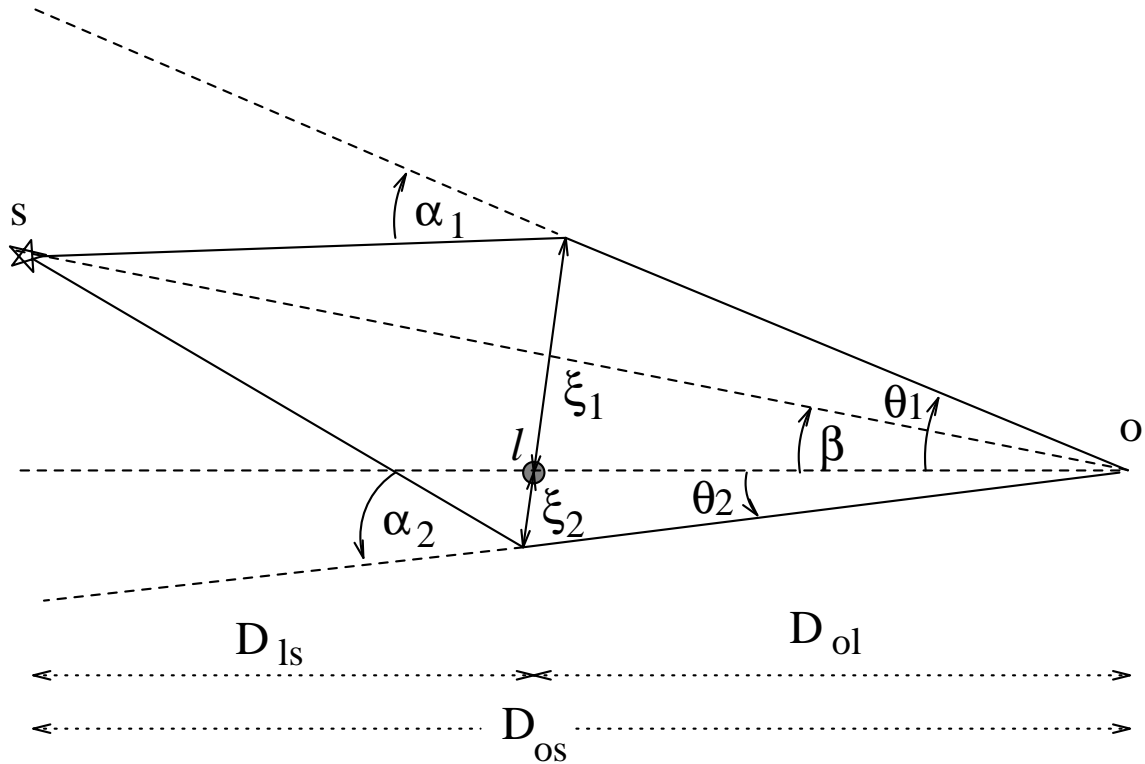,angle=90,height=10cm}}
\caption{Light ray trajectories in the plane containing the source
($s$), lens ($l$) and observer ($o$), and angular variables and
distances relevant for describing the ML event (see text).}
\label{angles.fig}
\end{figure}

When a compact object of mass $m$ comes close to the line of sight
(l.o.s.) to a background source star, it deflects its light, according
to general relativity\footnote{This prediction has been tested to 1\%
level with long baseline interferometry, looking for the change of the
apparent position of stars which come close to the limb of the Sun,
and also by the Hipparcos satellite, which measured light deflection
at impact parameters of $\sim 1$~AU with respect to the Sun, which are
comparable to those involved in ML searches.}, by an angle 
\begin{equation}
\alpha={4Gm\over c^2\xi}.
\label{alpha.eq}
\end{equation}
 Here $\xi$
is the impact parameter between the light ray trajectory and the
deflector, which we are assuming to be much larger than the 
Schwarschild radius $R_S\equiv2Gm/c^2$ of the deflector. 
This light deflection leads to
a lensing effect of the source light, and also to the appearance of a
secondary fainter image on the opposite side of the lensing object. The
geometry of the process in the plane containing the observer ($o$),
lens ($l$) and source star ($s$), is shown in
Fig.~\ref{angles.fig}.

Let $\theta_i$ ($i=1,2$) be the angles of the two source images with
respect to the observer--lens l.o.s., and $\beta$ be the angle to the
actual source position. Using that
$D_{os}(\theta_i-\beta)=D_{ls}\alpha_i$, with $\theta_i=\xi_i/D_{ol}$
and $\alpha_i=2R_S/\xi_i$, it is easy to obtain that
\begin{equation}
\theta_{1,2}={1\over 2}\left[\beta\pm\sqrt{\left({2R_E\over
D_{ol}}\right)^2+\beta^2}\;\right],
\end{equation}
where the angle $\theta_2$ to the fainter image is defined to be
negative. The Einstein radius 
\begin{equation}
R_E\equiv 2\sqrt{GmD_{ol}D_{ls}\over c^2D_{os}}
\label{re.eq}
\end{equation}
characterizes the size of the impact parameters for which the lensing
effect is large, and numerically turns out to be
\begin{equation}
R_E=9.0\ {\rm AU}\ \sqrt{{D_{os}\over 10\ {\rm kpc}}{m\over M_\odot}x(1-x)},
\end{equation}
where from here on we will often use the variable $x\equiv
D_{ol}/D_{os}$.

For perfect alignment, $\beta=0$, one has
$\theta_1=-\theta_2=R_E/D_{ol}$. Due to the symmetry of this
configuration, the image will actually be a ring, the so--called
Einstein ring, centered around the source position. Note that $R_E$ is
just the radius of the projection of this ring onto the lens plane (a
plane containing the lens and orthogonal to the l.o.s.).

In general, the angular separation between the two images is
\begin{equation}
\Delta\theta\equiv \theta_1-\theta_2=\sqrt{\left({2R_E\over
D_{ol}}\right)^2+\beta^2}.
\label{dtheta.eq}
\end{equation}
For impact parameters smaller than $R_E$ one then has 
\begin{equation}
\Delta\theta\simeq 1.8\sqrt{{(1-x)\over x}{10\ {\rm
kpc}\over D_{os}}{m\over M_\odot}}\ {\rm mas}. 
\end{equation}
This angle is then of
order 1~mas (milliarcsec) for ML observations of LMC stars, and
$O(\mu$as) for lenses and sources at cosmological distances
($\gg$~Mpc), as in the case of quasar ML by stars in foreground
galaxies.
Clearly these angular separations are unresolvable with present day
telescopes. 

\begin{figure}
\centerline{\psfig{figure=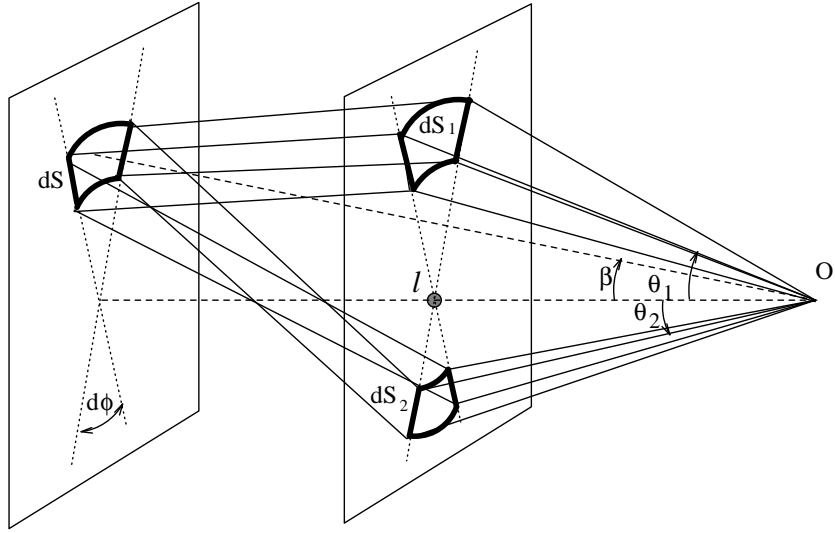,angle=90,height=7cm}}
\caption{Light ray trajectories from a surface element of the source
$dS$. The corresponding surface elements of the two images, projected
onto the lens plane, are $dS_1$ and $dS_2$.}
\label{solid.fig}
\end{figure}

The signal which can however be observed as a result of the ML process
is the variation of the intensity of the light source due to the
motion of the lensing object with respect to the l.o.s. to the source star.
A simple way to compute the resulting amplification of the light of
the source is to
compare the solid angles subtended by a surface element of the source
image ($d\Omega_i\equiv dS_i/(4\pi D_{ol}^2)$) 
with its value ($d\Omega_0\equiv dS/(4\pi D_{os}^2)$) in the absence of
lensing effects (see Fig.~\ref{solid.fig}). 
Since the source flux at the lens plane is constant
for the small impact parameters involved, the amplification of the
light intensity for each image, $A_i$, is just the ratio of these two 
solid angles. From Fig.~\ref{solid.fig} we obtain
\begin{equation}
A_i={d\Omega_i\over d\Omega_0}=\left|{\theta_id\phi d\theta_i\over \beta
d\phi d\beta}\right|.
\end{equation}

Substituting $d\theta_i/d\beta=(1\pm\beta/\Delta\theta)/2$ (where the
$+(-)$ sign is for $i=1(2)$) leads to a
total amplification
\begin{equation}
A\equiv A_1+A_2={1\over 2}\left[{\beta\over\Delta\theta}+{\Delta\theta
\over\beta}\right].
\end{equation}
Let us define $u\equiv\beta D_{ol}/R_E$ as the distance of the lens to
the l.o.s. to the star in units of $R_E$. We then obtain for the
amplification
\begin{equation}
A={u^2+2\over u\sqrt{u^2+4}}.
\label{amp.eq}
\end{equation}
Assuming that the lens moves with constant velocity with respect to
the l.o.s. during the duration of the ML event, one has 
\begin{equation}
u^2(t)={b^2+[v^\perp(t-t_0)]^2\over R_E^2}\equiv u_{min}^2+\left[{(t-t_0)\over
T}\right]^2,
\end{equation}
where $b$ is the minimum distance between the lens trajectory and the
l.o.s. to the star, $u_{min}\equiv b/R_E$, $t_0$ is the time of closest
approach and  $v^\perp$ is the component of the lens velocity relative
to the l.o.s. in the direction perpendicular to this same line.
The characteristic time 
\begin{equation}
T\equiv {R_E\over v^\perp},
\end{equation}
is now usually adopted as the definition of the event
duration\footnote{The MACHO group uses $\hat t\equiv 2T$ in their
analyses, while different event duration definitions \cite{gr91,de91} 
were used in the past but are less convenient.}.

\begin{figure}
\centerline{\psfig{figure=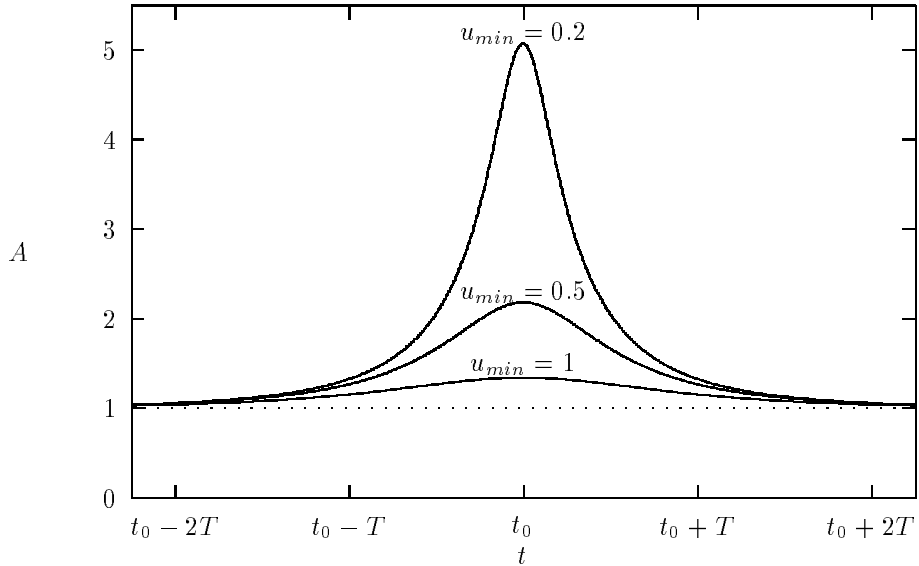,height=10cm}}
\caption{Light--curves, i.e. time evolution of the light amplification,
for different values of the normalised impact parameter $u_{min}$.}
\label{amp.fig}
\end{figure}

Combining the expression for $u(t)$ with Eq.~(\ref{amp.eq}) leads to a
time dependent amplification of the luminosity of the source described
by a very characteristic light--curve, plotted in Fig.~\ref{amp.fig}
for $u_{min}=1$, 0.5 and 0.2. The amplification reaches its maximum 
value $A_{max}$
at the time $t_0$ as $u(t)$ approaches $u_{min}$. In particular, we have
$A_{max}=1.34$ for $u_{min}=1$, and $A_{max}$ diverges (in the
point--like source approximation considered in this Section) for $u_{min}\to
0$. The curves are, of course, symmetric about $t_0$.

The event duration $T$ is determined, together with $u_{min}$ (or
equivalently $A_{max}$) and
$t_0$, by fitting the theoretical
light--curve to the observed star
luminosities plotted as a function of time.  There is clearly no useful
information contained in $t_0$, and the distribution of $A_{max}$
values tests only the expected uniform distribution  of impact parameters.  The
interesting information lies then in the event duration distribution
and in the rates of ML events. 
However, these observables depend in a convoluted
way on the lens and source spatial distributions and motions, as well
as on the lens mass function, so that it becomes very hard to extract
the lens properties in a clean way. 

Finally, a crucial characteristic  of the ML light--curve 
is that it should be achromatic, due to the gravitational origin
of the lensing effect which causes all wavelengths to be 
affected equally. This property, and the requirement of having an acceptable 
fit to the theoretical light--curve, 
are essential for discriminating the real ML events from other sources of
stellar variability\footnote{As will be discussed in Section~5, in
some cases the shape of the light--curve can deviate from Eq.~(\ref{amp.eq})
or not be achromatic, due to  binarity of the lens or source,
finite source size  or blending effects.}. 

\subsection{Optical depth}
A quantity which turns out to be useful in the discussion of ML is the
optical depth $\tau$ \cite{vi83}, which is the probability that at a given time a
source star is being microlensed with an amplification larger than
1.34, i.e. having  $u_{min}<1$.

If we assume for a moment that all the lenses  have the same mass $m$, this
probability is just the number of lenses inside a tube of radius $R_E$
around the l.o.s. to the star, assumed to be
at a distance $D_{os}$. The number of objects inside this
`microlensing tube' is 
\begin{equation}
\tau(D_{os})=\pi\int_0^{D_{os}}dD_{ol}{\rho_l\over
m}R_E^2={4\pi GD_{os}^2\over c^2}\int_0^1dx\ x(1-x)\rho_l(x).
\label{opt.eq}
\end{equation}
Since this expression turns out to be independent of the assumed 
lens mass, it actually holds also for any arbitrary lens mass distribution.

When performing observations in a given direction, one may have to
take into account, especially for bulge observations, that the sources
can be distributed with a non--negligible spread along the l.o.s.. 
 In this case, one
 has to perform an average over the source distribution \cite{ki94} to
obtain the actual optical depth
\begin{equation}
\tau={1\over N_s}\int dD_{os}{dn_s\over dD_{os}} \tau(D_{os}),
\label{tauav.eq}
\end{equation}
where the normalization factor is 
$N_s=\int dD_{os}dn_s/dD_{os}$ and $dn_s/dD_{os}$
describes the number density profile of detectable sources along the 
l.o.s.. For observations in the
bulge, it can be parameterised as 
$dn_s/dD_{os}\propto \rho_s D_{os}^{2-2\beta}$.
The factor $D_{os}^2$ is due to the change of the volume
element with distance. The parameter $\beta$ arises because the
observations are magnitude limited, and the
fraction of sources with luminosities larger than $L$ is assumed to
scale as $L^{-\beta}$, implying that the fraction of stars which
remain detectable behave as $D_{os}^{-2\beta}$. Hence, 
\begin{equation}
\tau={\int dD_{os}\ D_{os}^{2(1-\beta)}\rho_s\tau(D_{os})\over \int 
dD_{os}\ D_{os}^{2(1-\beta)}\rho_s}.
\label{optav.eq}
\end{equation}
A reasonable range for $\beta$ has been estimated to
be $\beta=1\pm 0.5$ in Baade's Window \cite{zh95}, and we will
generally use $\beta=1$ in the discussion of ML in the bulge.

\subsection{Event duration distribution}

Due to its simplicity and the property of being independent of the
lens motion and mass distribution, the optical depth is often used in
discussions of the ML observations. However, since the experiments
measure the number of events and their durations,
a more useful quantity is the
ML rate, and in particular its distribution as a function of the event
durations. This distribution contains all the information 
related to the ML process. 

\begin{figure}
\centerline{\psfig{figure=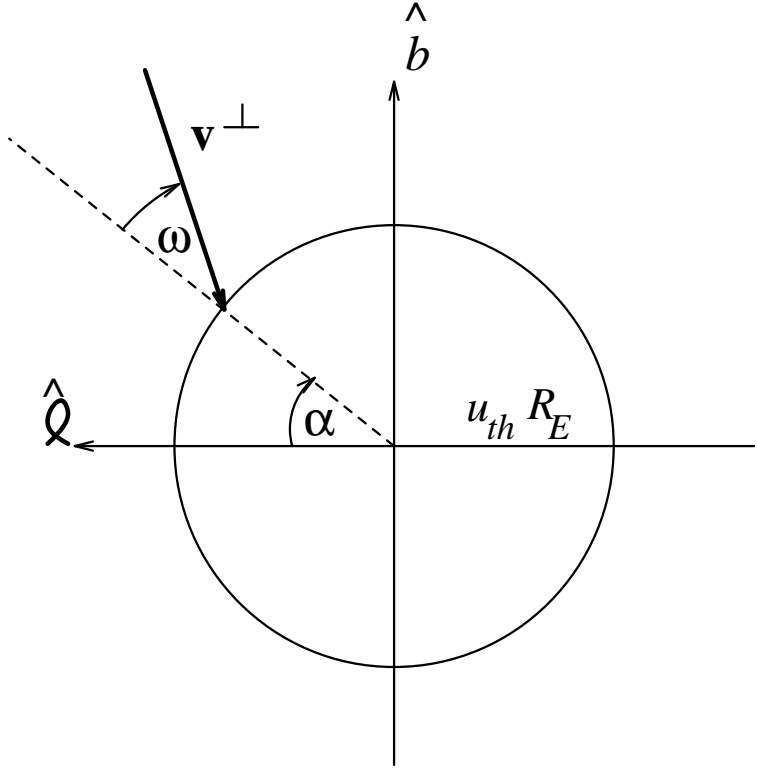,angle=90,height=10cm}}
\caption{ML tube section in a plane orthogonal to the
l.o.s. to the star. The tube  radius is $u_{th}R_E$, and we show the angular
variables entering the computation of the flux of lenses (moving with
relative velocity $v^\perp$ in this plane) through this tube.}
\label{mlt.fig}
\end{figure}

To obtain the event duration distribution, 
we start from the
expression for the differential rate in terms of the variables
depicted in Fig.~\ref{mlt.fig}. The rate of events with
amplification $A_{max}$ above a certain threshold $A_{th}$, i.e. with
$u_{min}<u_{th}$, is the flux of lenses inside a microlensing tube 
of radius $u_{th}R_E$ around the l.o.s. to the star. One then has
\begin{equation}
{d\Gamma\over dm}={dn\over dm}v^\perp\cos\omega f(
{\bf v}^\perp)d^2{\bf v}^\perp dS,
\label{dgam1.eq}
\end{equation}
where $dn/dm$ is the lens mass function, $dS=u_{th}R_Ed\alpha dD_{ol}$
is a surface element of the ML tube, $\omega$ is the angle between the
normal to the surface element and the lens relative velocity vector
${\bf v}^\perp$ (orthogonal to the l.o.s.). From here onwards we will
 adopt $u_{th}=1$, since it is enough to recall that 
the rates are just proportional to
$u_{th}$. The function $f$ in Eq.~(\ref{dgam1.eq})
describes the distribution of relative velocities. Due to the motion
of the observer and the source, with velocities ${\bf v}_o$ and ${\bf v}_s$
respectively,  the ML tube will be moving with a speed
\begin{equation}
{\bf v}_t=x{\bf v}_s+(1-x){\bf v}_o.
\end{equation}
If ${\bf v}_l$ is the lens velocity in the same reference frame, the
lens relative velocity with respect to the ML tube is
\begin{equation}
{\bf v}={\bf v}_l-{\bf v}_t.
\end{equation}
It is convenient to decompose all velocities as ${\bf v}_x={\bf
v}_{x,bulk}+{\bf v}_{x,dis}$, where ${\bf v}_{x,bulk}$ accounts for a bulk
global motion and ${\bf v}_{x,dis}$ is a dispersive contribution, which
will be assumed to be Gaussian distributed. From the previous
expressions we then have
\begin{eqnarray}
{\bf v}_{bulk}&=&{\bf v}_{l,bulk}-x{\bf v}_{s,bulk}-(1-x){\bf v}_o
\nonumber\\
{\bf v}_{dis}&=&{\bf v}_{l,dis}-x{\bf v}_{s,dis}.
\label{vel.eq}
\end{eqnarray}

The orthogonal component of ${\bf v}$ can be projected onto axes
along the directions of increasing longitudes, $\hat { \ell}$, 
and latitudes,
$\hat { b}$, i.e. ${\bf v}^\perp=v^\ell\hat { \ell}+
v^b\hat { b}$.
 The velocity distribution function can then be obtained from the
assumed Gaussian distribution $G(v^{i}_{dis})$ of the 
dispersive components ($i=\ell,b$)
\begin{equation}
G(v_{dis}^i)={1\over
\sqrt{2\pi}\sigma^i}{\rm exp}\left[-{(v_{dis}^i)^2\over
2(\sigma^i)^2} \right],
\label{gauss.eq}
\end{equation}
where $\sigma^i$ are the corresponding velocity dispersions. If
both lenses and sources have non--vanishing dispersions, $\sigma^i_l$
and $\sigma^i_s$ respectively, the relative velocity dispersion is just the
quadratic sum (see Eq.~(\ref{vel.eq}))
\begin{equation}
\sigma^i=\sqrt{(\sigma^i_l)^2+(x\sigma^i_s)^2}.
\label{sigi.eq}
\end{equation}
In terms of the modulus of the orthogonal relative velocity
$v^\perp$, which is the quantity related to the event duration
$T=R_E/v^\perp$, we can write (see Fig.~\ref{mlt.fig})
\begin{eqnarray}
v_{dis}^{\ell}&=&-v_{bulk}^\ell-v^\perp\cos\gamma \nonumber\\
v_{dis}^{ b}&=&-v_{bulk}^b-v^\perp\sin\gamma ,
\label{vdis.eq}
\end{eqnarray}
where $\gamma\equiv \omega+\alpha$.
We can now substitute in Eq.~(\ref{dgam1.eq}) 
\begin{equation}
f({\bf
v}^\perp)d^2{\bf v}^\perp=G(v_{dis}^\ell)G(v_{dis}^b)dv_{dis}^\ell d
v_{dis}^b,
\end{equation}
 and use that $d^2{\bf v}^\perp=v^\perp dv^\perp
d\omega$, to end up with
\begin{equation}
{d\Gamma\over dm}={dn\over dm} R_E(v^\perp)^2\cos\omega 
G(v_{dis}^\ell)G(v_{dis}^b)dv^\perp d\omega d\gamma dD_{ol}.
\end{equation}
The integral in $\omega\in [-\pi/2,\pi/2]$ (only lenses entering the
ML tube) is now trivial. 
The distribution in terms of the event duration $T=R_E/v^\perp$ is then 
\begin{equation}
{d\Gamma\over
dTdm}=2\int_0^{2\pi}d\gamma\int_0^{D_{os}}dD_{ol}{dn\over
dm}\left({R_E\over T}\right)^4G(v_{dis}^\ell)G(v_{dis}^b).
\label{dgdtdm.eq}
\end{equation}
The expression for the differential rate obtained here
is quite general, including an arbitrary lens mass
function, the effect of bulk velocities of observer, lens and
source\footnote{The bulk motion of the Sun is 
${\bf v}_\odot=(9,231,16)$~km/s (in the same coordinate system adopted
in Fig.~1). We
adopt for the LMC bulk motion ${\bf v}_{LMC}=(53,-160,162)$~km/s
\cite{gr91}. For the disk lensing populations there is a global
rotational motion
 while halo and spheroid populations are not expected to
have significant bulk motions.}
as well as the (eventually anisotropic) velocity dispersions 
of lenses and sources. In order to
apply it to a particular observational case, one needs just to
construct the appropiate $G(v^i_{dis})$ using Eqs.~(\ref{gauss.eq}),
(\ref{sigi.eq}) 
and (\ref{vdis.eq}), with the replacement $v^\perp=R_E/T$.  
The source
velocity dispersion is important for bulge observations, but can be
neglected for LMC stars lensed by Galactic objects. In this
last case, a further integral can be performed analytically if the
lens dispersion is isotropic \cite{gr91,gi94}. The effect of the bulk
motions is small for lensing of LMC stars by halo  objects \cite{gr91},
but becomes important  when considering the rotating disk
populations or for the discussion of bulge observations.

If the spread in source distances is non--negligible, an average over
the source locations similar to the one in Eq.~(\ref{optav.eq}) should
also be performed.

\subsection{Mass functions and time moments}
The determination of the lens mass function $dn(x)/dm$ is one of the
main purposes of ML observations. Very little is known about it
a priori. A usual assumption regarding its spatial dependence is the
so--called factorization hypothesis, i.e. that the mass function
 only changes with position by an overall normalization
\begin{equation}
{dn(x)\over dm}={\rho(x)\over \rho_0}{dn_0\over dm},
\end{equation}
where the subscript 0 stands for the local value of the lens
densities. We note that this may not be necessarily a good
approximation, since even for a single Galactic component, e.g. the
disk, the star formation conditions are  quite different in the high
density regions close to the bulge from those in the low density
regions in
the outskirts of the disk. One should also keep in mind that the mass
functions of stars in different Galactic components are not expected 
to be at all similar. 
In fact, already the measured spheroid star mass function
\cite{ri92,ri91} is completely
different from the disk one \cite{sc86}.

It is often useful to discuss ML
predictions under the simplifying assumption of having 
 a common lens mass $M$ as a first step, so that
\begin{equation}
{dn\over dm}={\rho\over m}\delta(m-M).
\end{equation}
In this case, one has then 
\begin{equation}
{d\Gamma\over
dT}={2\over M}\int_0^{2\pi}d\gamma\int_0^{D_{os}}dD_{ol}
\rho_l(D_{ol})(v^\perp)^4G(v_{dis}^\ell)G(v_{dis}^b).
\label{dgdtm.eq}
\end{equation}
Since $T$ only enters in this equation through $v^\perp\propto
\sqrt{M}/T$, 
 it is easy to see that the differential
rate distributions for two lens models with masses $M_1$ and $M_2$ are
related through
\begin{equation}
{d\Gamma\over dT}(T_1,M_1)={M_2\over M_1}{d\Gamma\over
dT}(T_2=T_1\sqrt{M_2/M_1},M_2).
\label{dgam12.eq}
\end{equation}
This implies that
the total `theoretical' rate 
\begin{equation}\Gamma_{th}=\int_0^\infty
dT {d\Gamma\over dT}
\end{equation}
 scales with the assumed lens mass as 
\begin{equation}
\Gamma_{th}
\propto 1/\sqrt{M}.
\label{gamth.eq}
\end{equation}

In order to actually predict the total event rate for a given
experiment, one has to take into account that the observational
efficiencies are generally smaller than unity and, moreover, that they
are  a function of the event duration,  mainly due
to the poor sampling of the short duration events and the incomplete coverage
of the very long duration ones. Hence, the predicted total rate for a
given experiment is a
convolution of the differential rate with the experimental 
efficiency $\epsilon(T)$, i.e.
\begin{equation}
\Gamma=\int_0^\infty dT\epsilon(T){d\Gamma\over dT}.
\label{gamma.eq}
\end{equation}
The efficiency $\epsilon(T)$ actually also takes into account the fact
that the observational threshold could be different from 
$A_{max}>1.34$, so that in Eq.~(\ref{gamma.eq}) the differential rate
should  be computed with $u_{th}=1$.

Other quantities which turn out to be useful are the expected time moments
\begin{equation}
\langle T^n\rangle ={1\over \Gamma}\int_0^\infty dT 
\epsilon (T){d\Gamma\over dT}T^n,
\end{equation}
where $n=1$ gives just the average event duration $\langle
T\rangle$. 

One may relate these time moments with the moments of the mass
distribution \cite{de91,je94b}, so that once the responsible lens
population is identified, its mass distribution may be reconstructed,
in the sense of knowing its moments, from the observed time moments. 
 A more elementary method for reconstructing the mass distribution 
is to use just an
 ansatz, depending on  few parameters, to describe a family of 
possible mass functions (e.g. power--law or Gaussian mass functions), 
 and use a likelihood method to determine
these parameters from a fit to the observed event durations
\cite{ha95,zh95,zh96}. 

A useful relation expressing the fact that 
the probability of ML $\tau$ is just proportional to the total 
rate  times the average event duration (computed with 
unit efficiency), is
\begin{equation}
\Gamma_{th}={2\over \pi}{\tau\over \langle T\rangle_{th}},
\label{gammatau.eq}
\end{equation}  
with the geometrical factor $2/\pi$ being related to the particular
definition of event duration adopted\footnote{One has
$\tau=\Gamma_{th}\langle t_e\rangle$, where $\langle t_e\rangle$ is the
average time during which $A>1.34$. Since $ T$ is the
 time it takes for the lens to move a distance $R_E$ orthogonaly to
the l.o.s., one has, for a given event,
$t_e=2T\sqrt{1-u_{min}^2}$. Making the average with a uniform
$u_{min}$ distribution leads to Eq.~(\ref{gammatau.eq}).}.

\begin{figure}
\centerline{\psfig{figure=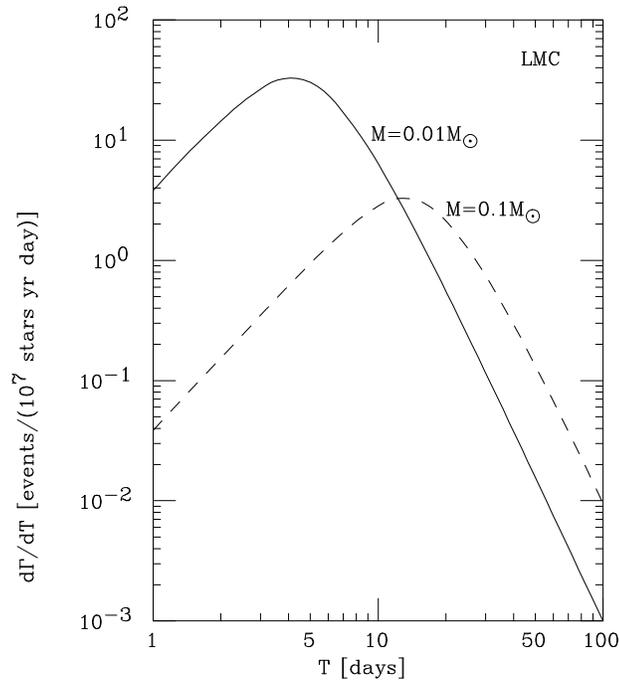,angle=90,height=10cm}}
\caption{Differential rate distribution for LMC stars lensed by a
standard halo population consisting of objects with a common mass of
$M=0.01M_\odot$ (solid line) and $M=0.1M_\odot$ (dashed line).}
\label{dgdt.fig}
\end{figure}

From this relation it is also clear that, in the case in which the
lenses are assumed to have a common mass $M$, the average event
duration scales with $M$ as
\begin{equation}
\langle T\rangle_{th}\propto \sqrt{M},
\label{tth.eq}
\end{equation}
just due to the fact that the associated Einstein radii scale as
$\sqrt{M}$. As an example of the expectations for the ML observations, 
Fig.~\ref{dgdt.fig} shows the predicted differential rate for LMC
stars lensed by objects in a standard Galactic halo, assuming two
different common lens masses, 
$10^{-2}M_\odot$ (solid lines) and $10^{-1}M_\odot$ (dashed lines).  
The behaviour mentioned in Eq.~(\ref{dgam12.eq}) 
is clearly apparent, and it is important to notice that the
distributions have a significant spread in the event durations, which
is even larger if one allows for a distribution in the lens masses
using a non--trivial mass function. This
 shows the difficulties underlying the determination of the
responsible lens masses from the observation of the event durations.

Another quantity useful for characterizing the theoretical models of
lensing populations is the average observer--lens distance, which is just 
\begin{equation}
\langle D_{ol}\rangle\equiv {1\over \Gamma_{th}}
\int_0^{D_{os}}dD_{ol}{d\Gamma\over dD_{ol}}D_{ol},
\end{equation}
eventually also averaged over the source spatial distribution.

\section{Microlensing searches}
In this Section we discuss the ongoing searches for ML events in
directions towards
the LMC and the bulge, and the possible interpretations of the first
results obtained.

\subsection{Microlensing signatures}

The main difficulty that ML searches have to face is the very small
probability of occurrence of events with observable amplifications. For
example, if the halo, distributed according to
 Eq.~(\ref{rhoh.eq}), wholly consists of compact objects
capable of producing ML events, the resulting optical depth for ML of LMC 
 stars is\footnote{From the virial theorem,
Eq.~(\ref{mhalo.eq}), together with Eq.~(\ref{rhoh.eq}) and using
$a<R_0\ll D_{os}$, it is easy to show that one expects
$\tau_{halo}\simeq (v_c/c)^2$.}
$\tau_{halo}\simeq
5\times 10^{-7}$. On the other hand, since the velocity dispersion of
halo objects is $\sigma\simeq v_c/\sqrt{2}$, an estimate of the typical
event duration is $T\simeq R_E/\sigma\simeq 100\ {\rm days}\sqrt{m/M_\odot}$
(taking $x\simeq 1/4$ in Eq.~(\ref{re.eq})). Hence, for lenses with
masses in the range $10^{-2}$--$10M_\odot$, the characteristic event
duration will be between a week and a year. Several 
million stars should then be  monitored for more than a  year
to get a handful of events in this case. For the lower end of the
interesting mass range, $10^{-7}$--$10^{-4}M_\odot$, the events last
less than a day, and the event rates, $\Gamma\propto\tau/\langle
T\rangle$, are much larger. 
To be sensitive to this mass range, fewer stars need to be followed but
with a much better time coverage, i.e. performing several measurements per
night. Both strategies have actually been employed
 in order to explore the whole range 
 $10^{-7}<m/M_\odot<1$. 
The computation of the optical depth for bulge stars, lensed by low
mass stars in the bulge  and disk, also leads to $\tau\simeq
10^{-6}$, with typical event durations of a few weeks (see below), so that
in this case also several million stars need to be monitored for more
than a year to get significant statistics.

In addition to the small number of expected events, a further
complication is that there is a large background of variable
stars, and special care should be taken
in order to avoid  including one of these stars as an ML event.
Variable stars represent a fraction $\sim 3\times 10^{-3}$ of the total
number of stars, and  
 an interesting byproduct of the ML searches has been
the compilation of the 
more complete catalogs of variables in the LMC \cite{al95f,gr95b,be95c}
and in Baade's window \cite{ud94f}.

There are many signatures which allow one to distinguish between  intrinsic
stellar variabilities and  ML events, so that one can be confident
that a real signal of ML is found:
\begin{itemize}
\item Most variables are periodic, while  due to the very small
probability for a star to be lensed, ML events should never repeat
for the same star.
\item Most intrinsic luminosity variations have an associated
change in the star temperature, so that the variation is color
dependent. ML events are instead achromatic, due to their
gravitational nature (unless there is blending of two sources with
significantly different colors).
\item Most variable stars produce pulses which are asymmetric in
time, with a rapid rise and a slower decline of the star luminosity,
while ML events are time symmetric.
\item The light--curves of real ML events should be reasonably
fitted by the theoretical light--curve, Eq.~(\ref{amp.eq}), eventually
allowing for stellar blending or for peculiarities such as lens or
source binarity, finite source size or the non--uniform motion of the Earth
 (see  Section 5).
\end{itemize}

\begin{figure}
\centerline{\psfig{figure=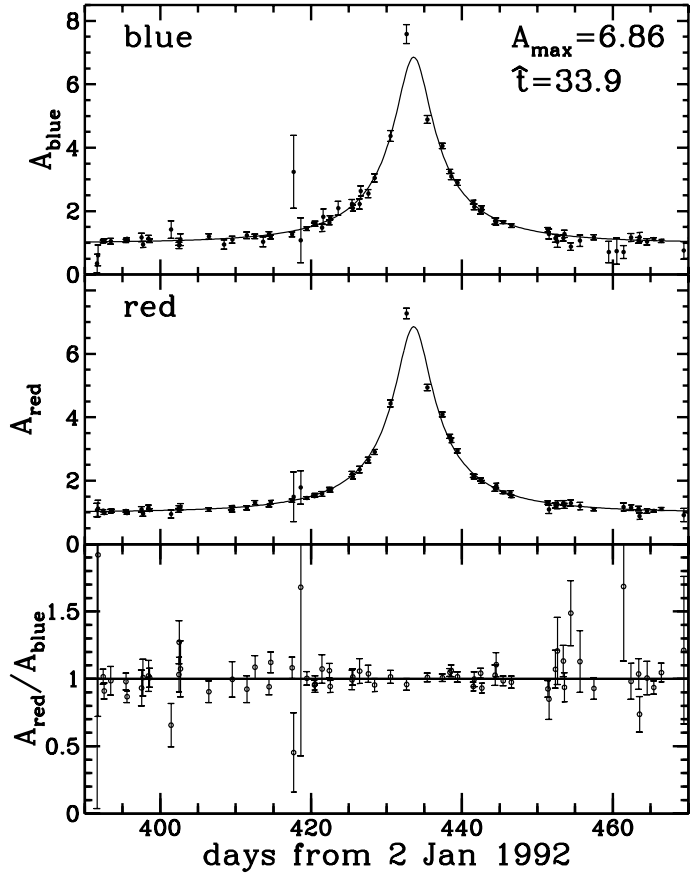,height=12cm}}
\caption{Gold--plated event obtained by the MACHO collaboration in
their LMC analysis.}
\label{gold.fig}
\end{figure}

These general features can be seen for instance in
Fig.~\ref{gold.fig}, which shows the light--curve of the first
ML event observed by the MACHO collaboration in the LMC\footnote{A
more recent fit to the observed amplifications of this event gives
$A_{max}=7.2$ \cite{al95e}.} (from Ref.~\cite{al93}). The
light--curves in the two colors are reasonably well fitted by the
theoretical curve, and the ratio of the amplifications in each color
(lower panel) is well consistent with an achromatic signal. Accurate
 testing of the achromaticity, and even of the
stability of the source spectrum \cite{be95a}, during the ML event is
now possible with the Early Warning systems implemented both by the
MACHO \cite{pr95} and OGLE \cite{ud94d} collaborations. This allows
one to
know the existence of an ongoing ML event in real time, as the
amplification is starting to grow, and hence to study the 
 source properties with other bigger telescopes and to organize intense
follow--up studies of the light--curves with telescope networks around
the globe \cite{pr95}.

In addition to the above mentioned features, 
ML events should have the following statistical properties:
\begin{itemize}
\item Unlike star variability, ML events should happen with the
same probability for any kind of star, and this should reflect in the
distribution of ML events in the color magnitude
diagrams\footnote{For observations in the bulge, however, since source
stars have a non--negligible spread along the l.o.s. and $\tau$
is significantly larger for the stars lying behind the bulge, the
lensing probabilities should increase for fainter stars. On the other
hand, these probabilities should be negligible for foreground disk
stars.}.
\item The distribution of $A_{max}$ values should be the one
corresponding to a uniform distribution of $u_{min}$. 
\item The distribution of $A_{max}$ and $T$ values should be
uncorrelated (once the effects of blending are taken into account
\cite{al95e}). 
\end{itemize}

Due to the large number of stars that need to be followed to observe
ML events, the ongoing searches have focused on two
targets: $i)$ Stars in the Large and
Small Magellanic Clouds, which are the nearest galaxies having
l.o.s. which go out of the Galactic plane and well across the halo. 
 $ii)$ Stars in the Galactic bulge, which allow one to test the
distribution of lenses near to the plane of the Galaxy.

Globular Clusters, having less than $10^5$ stars, could only be useful
for testing lens masses below Jupiter's mass, although it has been pointed
out that they could provide an interesting lensing population, using
some clusters which are in the front of the bulge or SMC fields
\cite{pa94c,ta95}. On the other hand, observations of  ML of
stars in the Andromeda galaxy (M31) are starting to be performed, but
since M31 is much further away ($D_{os}\simeq 770$~kpc), the severe
crowding of stars requires the use of `pixel' lensing, since the
individual stars cannot be resolved (see Section 6).

\subsection{Searches towards the Magellanic Clouds}

\subsubsection{Experimental results}

Two experiments are now looking for ML in the Magellanic Clouds. 
The first is the
French EROS (Exp\'erience de Recherche d'Objets
Sombres) experiment\footnote{The EROS home--page is at
http://www.lal.in2p3.fr/EROS/eros.html}, which actually had two
different programs at La Silla Observatory in Chile:
$i)$ they have used a CCD camera in a 40~cm
dedicated telescope to make short exposures (10~min each) so as to be
able to test short duration events; $ii)$
 they have analysed plates
from a 1~m Schmidt telescope, which made two exposures per night in
different colors, from which they have followed the light--curves of
several million stars since 1991. They have also one year of data from
the SMC yet to be analysed. 
From 1996, they 
plan to start observations with a new dedicated 1~m telescope,
equipped with two 
CCD cameras, and this will improve their statistics significantly.  

The second is the MACHO (Massive Compact Halo Objects)  experiment
\footnote{The MACHO home--page is at http://wwwmacho.anu.edu.au}, 
an Australian/American collaboration using a 1.3~m dedicated telescope
at the Mount Stromlo Observatory in Australia. They can 
simultaneously take images in a `red' and a `blue' band using two CCD
cameras. They have measured the light--curves of $\sim 10^7$ stars in
the LMC since 1992, and they also have measurements in the SMC, though
with poorer statistics.

The EROS group has found no events in their CCD search, after looking
at the light curves of 82000 stars during 10 months \cite{au95}. 
In the first
three years of Schmidt plate data \cite{an96}, 
with a total exposure $E=3\ {\rm
yr}\times 3.33\ 10^6$~stars, they found two candidate ML events,
with durations $T_1=23$~d and $T_2=29$~days\footnote{The 
source star of the second event
was later found to be an eclipsing binary, with period of 2.8~d,
making the interpretation of the event as due to ML less reliable
\cite{an95}.}. From the two events found, and their
efficiency $\epsilon(T)$, they estimate an optical depth
towards the LMC of 
\begin{equation}
\tau^{EROS}_{est}\equiv {\pi\over 2E}\sum_{events}{T_i\over
\epsilon(T_i)}= 8.2\times 10^{-8}.
\label{taueros.eq} 
\end{equation}

The analysis of the first year of MACHO observations \cite{al95e}, using 9.5
million light--curves, revealed the presence of three ML events, with
durations $T=19.4$, 10.1 and 15.6~d (see Note added).  Performing a 
likelihood analysis, they inferred from these events an 
 optical depth towards the LMC  
\begin{equation}
\tau^{MACHO}=8.8^{+7}_{-5}\times 10^{-8}.
\label{taumacho.eq}
\end{equation}
An estimate of the optical depth as in Eq.~(\ref{taueros.eq}) would
have lead to $\tau_{est}^{MACHO}=8\times 10^{-8}$, in good agreement
with the maximum likelihood estimate. 

The values of $\tau$, Eqs.~(\ref{taueros.eq}) and (\ref{taumacho.eq}),
should however be taken cum grano salis since, for instance, lenses
with masses larger than 0.1~$M_\odot$, to which the experiments are not
very sensitive yet, could in principle contribute significantly to $\tau$. On
the other hand, if some of the events are not due to real ML, the
optical depth could be smaller. Also, it has been noted
\cite{ha95c} that the error associated with $\tau$ is larger than the
naive Poissonian estimate, due to the wide range of expected
timescales. As a result, it is always convenient to confront directly the
expected event rates and durations, although this  requires particular
assumptions concerning the lens mass function.

\begin{table}
\begin{center}
\begin{tabular}{c|cccc} \hline\hline
  Lensing  & $\tau$ & $\Gamma_{th}$ & $\langle T\rangle_{th}$ & 
$\langle D_{ol}\rangle$ \\
component & [$10^{-7}$] &  [$(10^7$ stars yr)$^{-1}$] & [days] &[kpc]\\
\hline
Halo&5.4 & $89/\sqrt{m_{0.05}}$ & $14\sqrt{m_{0.05}}$&14   \\
Spheroid&0.44$\rho^S_*$&$6.7\rho^S_*/\sqrt{m_{0.05}}$&
$15\sqrt{m_{0.05}}$&9.0   \\
Thick disk&$0.47\Sigma_{45}$&
$5.5\Sigma_{45}/\sqrt{m_{0.05}}$&$20\sqrt{m_{0.05}}$ & 3.6\\
Thin disk&$0.11\Sigma_{30}$& $1.1\Sigma_{30}/\sqrt{m_{0.05}}$&
$23\sqrt{m_{0.05}}$& 1.1\\
\hline
\end{tabular}
\end{center}
\caption{Predicted optical depth $\tau$, theoretical ($\epsilon=1$) rates 
$\Gamma_{th}$,  average event duration
$\langle T\rangle_{th}$ and average lens distance 
$\langle D_{ol}\rangle$, for ML
of LMC stars by different Galactic 
lensing populations: a standard halo, a heavy
spheroid (where $\rho^S_*\equiv\rho^S_0/0.0015M_\odot$~pc$^{-3}$), 
 a dark thick disk (with
$\Sigma_{45}\equiv\Sigma_0/(45M_\odot$~pc$^{-2}$) and thin disk  (with
$\Sigma_{30}\equiv\Sigma_0/(30M_\odot$~pc$^{-2}$).}
\end{table}

In Table~1 we give the values for $\tau$, $\Gamma_{th}$, $\langle
T\rangle_{th}$ 
and $\langle D_{ol}\rangle$ for observations towards the LMC, for 
the different dark lensing  components discussed in
Section~2. The theoretical rates and average durations in the Table
are  computed assuming a common lens mass and unit efficiency, while
the actual efficiencies of the experiments are typically $\leq 30$\%
and $T$ dependent \cite{al95e,an96}. In any case, taking into account
the efficiencies would lead to more than 10 expected events in any of the
searches sensitive to long duration events, arising from a
standard halo model\footnote{The standard halos used by the EROS and MACHO
groups are slightly different, and also slightly lighter than the one
used in Table~1.} wholly consisting of objects with masses between
$10^{-4}$ and $10^{-1}M_\odot$ \cite{an96,al95e}. 
In the EROS CCD search, more than 2.3 events (the 90\% CL upper bound
corresponding to no observed events) would be expected from a standard
halo composed of objects with $5\times 10^{-8}<m/M_\odot<7\times 10^{-4}$
\cite{au95}. 

\begin{figure}
\centerline{\psfig{figure=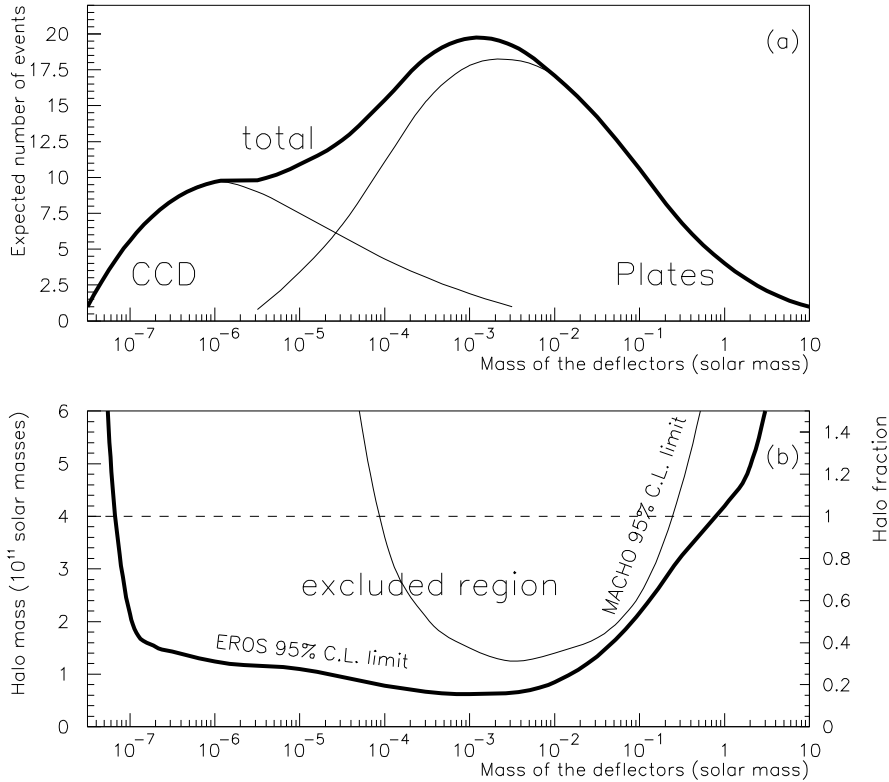,height=12cm}}
\caption{LMC bounds from EROS results. The upper panel shows
the expected number of events  for the two EROS searches 
from a standard halo wholly composed of
compact objects as a function of their assumed common mass. 
The lower panel
shows the corresponding 95\% CL bounds on the halo mass contained
within 50~kpc, or alternatively (right scale) the allowed fraction of
compact objects in a standard halo. Also the bounds from the first
year LMC MACHO results are shown.}
\label{lmcbound.fig}
\end{figure}

From these results it is possible to set an upper bound
on the contribution to the halo mass coming from objects with masses
in the range $5\times 10^{-8}<m/M_\odot<1$, which is shown in
Fig.~\ref{lmcbound.fig}, taken from Ref.~\cite{an96}. 
The upper panel shows the expected number of
events in the two EROS experiments for halo models in which all lenses
are assumed to have the same mass. The lower panel gives the resulting
95\% CL  constraints on the halo mass within 50~kpc, 
 or alternatively (right scale) the fraction $f$ of the mass
of a standard halo which is allowed to consist of
 compact objects. The thick line is the combined
EROS result, while the thin line is the MACHO excluded region. From
this we see that in the wide range $10^{-7}<m/M_\odot<10^{-1}$, the
halo fraction should be less than 30\%. This last result is also
independent of the assumption that the lenses have a common 
mass, as long as all the masses are inside that range.
 Although the value of this fraction depends strongly on the
reference halo model adopted, the bound on the halo mass (left scale)
is quite insensitive to this, providing then a more solid result
\cite{al95,al95e,ga95c}.

It is also possible to estimate the lens masses which are more likely
to produce the observed event durations. Since the durations observed
 actually have a narrow dispersion, it is reasonable to compare with
models having a common lens mass, since an extended mass function would
only increase the spread in the expected event durations. From a
maximum likelihood fit, the MACHO group \cite{al95e} 
finds that for the standard
halo, the more likely lens masses turn out to be in the range
$m=0.065^{+0.06}_{-0.03}M_\odot$ (68\% CL), 
which correspond to brown dwarfs or, in
the upper side, to very faint stars.    The EROS group
\cite{an96}, which  observed somewhat larger durations, infer a `95\% CL'
interval [0.01--0.7$]M_\odot$ from the $T=23$~d event, while an
allowed mass interval of
[0.02--1.1$]M_\odot$ from the $T=29$~d event.

\subsubsection {Interpreting the observations}

The results discussed in the last Section indicate that the observed
ML rates and optical depths towards the LMC are significantly 
smaller than those
expected from a standard halo consisting entirely of compact
objects with masses in the range $10^{-7}$--$10^{-1}M_\odot$ (see Note
added).  However, they are larger than those that would result from 
the known populations of faint stars in the Galactic disk, which lead
to  $\tau\sim 10^{-8}$ \cite{go94c},
 and from those in the LMC itself, which lead to a prediction 
$\tau\leq 3\times 10^{-8}$ \cite{wu94,de95b,al95e}.

Several possible scenarios have been suggested to account for these
observations, which we now discuss:

\begin{itemize}
\item The halo consists of a fraction $f\simeq0.1$--0.3 of
compact baryonic objects \cite{al95e,al95b,an96,ga94,ga95,ke95}, 
with the remaining 70--90\% being either in cold baryonic gas
\cite{pf94,de95,ge95}, which
does not produce any lensing effect (see however \cite{he95}), or in
white dwarfs and neutron star stellar remnants or 
heavy  black holes, to which the present searches are
almost insensitive, or alternatively in non--baryonic forms.

\item The halo may actually be much lighter than the
reference models usually considered. Varying the halo models so as to
allow for declining rotation curves or a maximum contribution to $v_c$
from the Galactic disk, can result in models with much larger allowed
mass fraction in compact objects, $f\geq 0.5$ 
\cite{ev94b,al95,al95e,ga95c,ka95b}. Fractions $f\simeq 1$ may even be
consistent with the ML results, although they seem disfavoured when
combined with other Galactic dynamical constraints
\cite{ga95c}. Allowing for
anisotropic velocity distributions can further affect, in a sizeable
way, the rates
and event durations \cite{de95c,ev95b}. For instance, 
a tangentially anisotropic velocity distribution leads to a rate which
is a factor of two larger than the one resulting from a radially
anisotropic distribution, with the isotropic distribution usually
adopted giving an intermediate value.
Allowing for halo flattening does not modify sizeably
the predictions for LMC observations \cite{sa93}, due to a
compensation between the increase in the local halo density required
to keep unchanged the rotation velocity and the fact that the lenses
become closer on average.  
 Tilting
a flattened halo may however have some effects \cite{fr94}. The effects of
prolate halos were discussed in Ref.~\cite{ho95}, and they may reduce the
optical depth by up to 35\%, although prolate shapes seem unnatural. 

One must also mention that some observations suggest that  the LMC has 
a dark halo \cite{we91,sc92b} and, of course, if the Milky Way halo
consists of compact baryonic objects the LMC one should also consist
of similar objects. In this case the 
contribution of the LMC halo to the optical depth is $\sim
20$\% of that of a standard Galactic halo \cite{go93,wu94,de95b}, and hence
is already comparable with the observed depth. Its
inclusion would then further increase the gap between the observations
and the theoretical halo model predictions.

\item The halo may be completely non--baryonic,
i.e. $f=0$, with the
observed events being due to dark objects in the Galactic spheroid
\cite{gi94} or in the Galactic thick disk \cite{go94,go94c}. From
Table~1 we see that, if the density of the outer--spheroid is similar
to that of the heavy spheroid models \cite{ca81,os82}, or if the
local column density of the thick disk is close to its upper bound
$\Sigma\simeq 45M_\odot/$pc$^3$ (see Section 2), these models give
results which are consistent with the observations. The inferred lens
masses for these models are similar to those inferred for the halo
case, since the slower motions are compensated by the fact that the
lenses are typically closer, having then smaller associated Einstein
radii \cite{go94c,de95b}.
The additional  contribution arising from  the faint stars in
the Galactic thin disk and in the LMC itself plays a
non--negligible role in these scenarios.

\item It has been proposed \cite{sa94} that the contribution from
LMC stars may be enough to explain the observations. In
Ref.~\cite{sa94}, an optical depth $\tau\simeq 5\times 10^{-8}$ is
obtained for this contribution,
assuming that the LMC bar is quite massive. Other estimates of the optical
depth of stellar LMC populations lead however to smaller values
\cite{wu94,go95d,de95b}, and the estimated event durations for lensing
by faint stars in the LMC are somewhat larger than the observed ones
\cite{de95b,al95e}. Another difficulty with this interpretation is that
the two EROS events are actually outside the LMC bar, in a region
where the LMC stars should contribute negligibly to the optical depth,
but it could still be allowed  if one
considers that some of the events may not actually be due to ML.
\end{itemize}

In order to distinguish among these alternative scenarios, clearly more
statistics are required. The main tests that will clarify the
situation are the following:

\begin{itemize}
\item If the lenses are Galactic, rather than in the LMC,
the optical
depth should be independent of the location of the source star in
the LMC. Hence, a strong variation of $\tau$ across the LMC will be a
clear signal that the lenses are in the LMC itself.

\item The Einstein radius of a lens
 projected onto the source plane is 
\begin{equation}
{R_E\over x}\simeq4.3\times 10^3R_\odot\sqrt{\left({m\over
M_\odot}\right)\left({1-x\over x}\right)} .
\end{equation}
Hence, if the lens is a star in the LMC, 
i.e. $1-x\leq 10^{-2}$ and $m\geq 0.1M_\odot$, this quantity is
comparable with the size of a red giant star. It then becomes more
likely to observe deviations in the light--curve due to the finite
source size (see  Section 5). For Galactic lenses, the projected
radius is much larger, making  the observation of
 such effects very unlikely.
\end{itemize}
To discriminate between Galactic lenses in the thick disk or in the
 spheroid looks
particularly difficult, since they lead to similar predictions for the
LMC rates and average event durations (see Table 1). Some strategies
have been proposed to study this issue:
\begin{itemize}
\item
 If sufficient data are gathered from the SMC, the ratio between
the rates obtained for the SMC and the LMC should be quite different
in the two scenarios. Since the SMC is at a smaller angle with respect to
the Galactic center than the LMC, but however at larger latitudes, one
has $\Gamma_{SMC}/\Gamma_{LMC}\simeq 1.7$ for the spheroid lenses,
while $\Gamma_{SMC}/\Gamma_{LMC}\simeq 0.9$ for the thick disk
\cite{de95b}.  Of course, this assumes a negligible contribution
from lensing populations in the Magellanic Clouds 
themselves. The comparison of LMC and SMC rates was
originally proposed as a tool to study the ellipticity of the Galactic
halo \cite{sa93}, and then suggested as a way to distinguish between
the thick disk and the halo \cite{go94c}.
\item Since thick disk objects have small velocity
dispersion ($\sim 40$~km/s) and a global rotational motion, thick
disk lenses cross the l.o.s. to LMC stars at an approximately
constant speed (for a given value of $x$). This implies that  
the event duration distribution due to thick disk lenses has a smaller
dispersion, $\Delta T\equiv\sqrt{\langle T^2\rangle-\langle
T\rangle^2}/\langle T\rangle$, than the 
 spheroid or halo ones. This difference is still present
if one takes into account the spread in $T$ due to the ignorance of
the lens mass function, as long as lens masses heavier than the Jeans
mass $\simeq 7\times 10^{-3}M_\odot$ are considered \cite{de95b}.
Hence, with several dozens of events this test may become useful.
\item
Lenses belonging to the Galactic  disk populations, which are on
average closer to the observer (see Table 1), may lead to parallax
effects due to the motion of the Earth around the Sun (see
Section 5). This parallax may be observable for most of the thin disk
events, and for around 15\% of the thick disk ones \cite{go94c}. 
\end{itemize}
\subsection{Searches towards the bulge}

\subsubsection{Experimental results}

Three experiments are looking for ML towards the Galactic center, and
they have obtained a large number of ML events, well above the initial
expectations. The OGLE (Optical
Gravitational Lensing Experiment) experiment\footnote{The OGLE
home--page is at http://www.astrouw.edu.pl} is a collaboration between 
Warsaw and the Carnegie Institute, using a 1~m telescope with a CCD
camera at Las Campanas Observatory, Chile. Its main purpose is the study
of ML and variable stars in the Galactic bulge, where they have
monitored $\sim 10^6$ stars since 1992. Most of the exposures are
taken in the $I$ band, and fewer $V$ band pictures are taken to check the
achromaticity. 
The MACHO collaboration also looks at fields in the
bulge, especially  when the LMC is low in the sky. They have followed
more than $10^7$~stars since 1993.  The DUO (Disk Unseen Objects)
experiment is a French collaboration which has analysed Schmidt plates
taken (two in red and one in blue per night) at the ESO 1~m telescope 
at La Silla, Chile, since 1994 \cite{al95a}. They have found a dozen
events, but the analysis is not yet available.

The analysis of the first two years of OGLE data \cite{ud94b} 
provided 9 events,
with 8.6~d$<T<62$~d, including a binary lens candidate, 
out of $10^6$ light--curves of stars in Baade's Window and
in other fields at $\ell=\pm 5^\circ$ and $b=-3.5^\circ$. The
distribution of peak amplifications is consistent with ML expectations
and the lensed stars are scattered in the color--magnitude diagrams
as if taken at random from the overall distribution of bulge stars.
From these results they estimate an average optical depth for bulge
sources of
\begin{equation}
\tau^{OGLE}=(3.3\pm 1.2)\times 10^{-6}.
\label{tauogle}
\end{equation}

\begin{figure}
\centerline{\psfig{figure=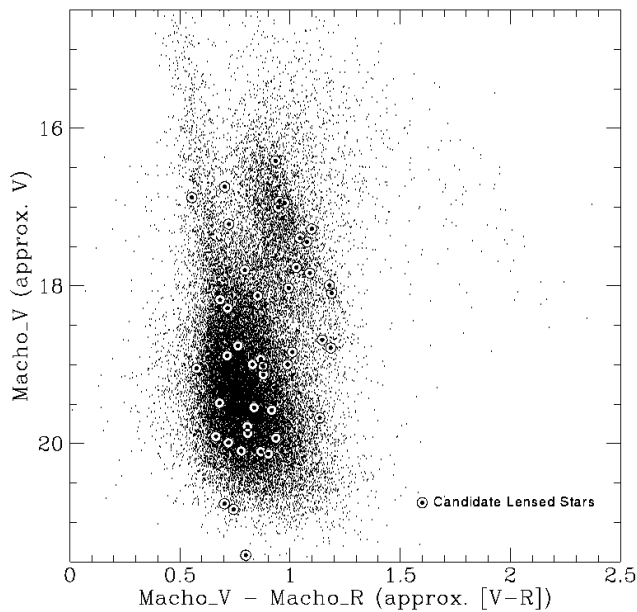,height=10cm}}
\caption{Color--Magnitude diagram of a sample of the stars monitored
in the bulge by the MACHO collaboration showing with
circles the location of the first 45 bulge ML events, 
reprinted with permission from The Astrophysical Journal.}
\label{cmd.fig}
\end{figure}

The searches of the MACHO collaboration have returned 45 ML
events towards the bulge in their first year, looking at
several fields with $0^\circ<\ell<7^\circ$ and
$-2^\circ>b>-6^\circ$ \cite{al95c,al96},
 and there are already more than 40 events seen in
the 1995 season. The observed event durations range from 4.5~d
to 110~d. An analysis based on 13 events detected  among
$1.3\times 10^6$ clump giant stars\footnote{These are low mass He core
burning giants.}, in fields centered at $\ell=2.5^\circ$ and
$b=-3.6^\circ$, leads to an estimated average optical depth
\cite{al96}
\begin{equation}
\tau^{MACHO}=3.9^{+1.8}_{-1.2}\times 10^{-6}.
\label{taumacho}
\end{equation}

The color--magnitude diagram for the first 45 candidate ML events
(from Ref.~\cite{al96}) is
shown in Fig.~\ref{cmd.fig}. The red clump stars are centered at
$V\simeq17$ and $V-R\simeq 1$. The events are distributed all over the
diagram. The comparatively large fraction of lensed stars for $V<19$
is partly due to the larger ML efficiency associated with the brightest
stars, although there are suggestions \cite{al96} that the optical depth is
smaller on the average than the result in Eq.~(\ref{taumacho})
associated with the red clumps. This could be related to the fact
that a fraction of the sources are foreground disk stars, having then
smaller associated optical depth, and this fraction is minimized in
the red clump subsample. Also, since red clumps are very bright stars,
they can be seen further away than main sequence stars ($\beta\simeq
0$ in Eq.~(\ref{tauav.eq})) and hence are more likely to be lensed.
The effect of disk lenses and sources should
be enhanced at smaller latitudes and at larger longitudes, hence the
relevance of gathering events in several fields so as to be able to
map the ML rates in different directions and from this infer which are
the populations responsible for the observations. There are, in fact,
preliminary indications that the optical depth increases at smaller
latitudes \cite{al96}. 

\subsubsection{Interpreting the observations}

The ML searches towards the bulge were initially believed to be
sensitive mainly to disk lenses, and in this sense these observations,
unlike the LMC ones, were warranted to provide events independently of
the existence of compact objects in the halo. The optical depth for
bulge stars lensed by faint disk stars was expected to be
$\leq10^{-6}$ \cite{gr91b,pa91}, as shown in the row labeled `faint
stars' in Table 2, which was computed by Griest et al. 
\cite{gr91b} assuming a fixed
local number density of disk stars and allowing their mass function to
vary consistently with the errors in its determination. The
contribution of the dark halo was not expected to be dominant,
although the predictions depend sensitively on the assumed halo core
radius $a$, increasing with decreasing $a$, and they also increase
with  the halo
flattening \cite{ga95c}. 
It was then shown \cite{gi94} that in the heavy spheroid
models the contribution of spheroid lenses  was comparable to that
of disk lenses, and arose mainly from the inner 2~kpc of the Galaxy
\cite{ro94}, where this component describes the bulge. However, the
associated event durations were typically smaller than the disk ones. 
Kiraga and
Paczy\'nski \cite{ki94} studied the bulge--bulge ML with Kent's
axisymmetric model, showing also the importance of averaging over the
source distribution (Eq.~(\ref{optav.eq})), since this increases both the
bulge--bulge rates and the average event durations  by $\sim 25$\%
(increasing $\tau$ by $\sim 50$\%). This averaging however does not
significantly change the rates due to disk lenses, because these 
are on average more distant  from the sources and less centrally
concentrated than the bulge lenses. They also considered how the rates
due to disk lenses are reduced in models where a `hole' in the
central distribution of disk stars is present (see Section~2.4). To
illustrate this last point, 
we show in Table~2 the predictions for a disk with a
hole in the inner 2.5~kpc (i.e. considering only $D_{ol}<6$~kpc), as
well as those for a disk without a  hole. Finally, they also discussed
the dependence of the predictions on the parameter $\beta$
 (see Eq.~(\ref{tauav.eq})) 
describing the change with distance of the fraction of sources which
remain observable. 

\begin{table}
\begin{center}
\begin{tabular}{c|cccc} \hline\hline
  Lensing  & $\tau$ & $\Gamma_{th}$ & $\langle T\rangle_{th}$ & 
$\langle D_{os}\rangle$ \\
component & [$10^{-6}$] &  [$(10^6$ stars yr)$^{-1}$] & [days] &[kpc]\\
\hline
Faint stars& 0.29--0.96& 2.2--7.5&$\sim 30$&5.7\\
Bar($\alpha=10^\circ$)&$1.74\ M_{2}$& $19\ M_2/\sqrt{m_{0.2}}$&
$20\sqrt{m_{0.2}}$ & 7.0\\
Bar$(\alpha=30^\circ$)&$0.97\ M_{2}$ & $12\ M_2/\sqrt{m_{0.2}}$&
$18\sqrt{m_{0.2}}$ & 7.4\\
Disk($D_{ol}<12$)&$0.42\ \Sigma_{30}$& $4.3\ 
\Sigma_{30}/\sqrt{m_{0.2}}$ & $22\sqrt{m_{0.2}}$ & 5.5\\
Disk($D_{ol}<6$)&$0.31\ \Sigma_{30}$& $2.5\ 
\Sigma_{30}/\sqrt{m_{0.2}}$ & $28\sqrt{m_{0.2}}$ & 4.1\\
Thick disk&$0.37\ \Sigma_{45}$&
$3.3\ \Sigma_{45}/\sqrt{m_{0.2}}$& $25\sqrt{m_{0.2}}$ & 6.2\\
Spheroid&$0.72\rho^S_*$& $14\rho^S_*/\sqrt{m_{0.2}}$&$12\sqrt{m_{0.2}}$ &7.7 \\
Halo&0.24& $4.7/\sqrt{m_{0.2}}$ &$12\sqrt{m_{0.2}}$& 5.4\\
\hline
\end{tabular}
\end{center}
\caption{Predicted optical depth $\tau$, `theoretical' rates 
$\Gamma_{th}$, average event duration
$\langle T\rangle_{th}$ and lens distance $\langle D_{ol}\rangle$, for ML
of stars in Baade's window by different lens populations (see text and
caption of Table~1). For the disk and halo predictions we assumed the
sources to be in the bar.}
\end{table}

Since the observed optical depth, Eqs.~(\ref{tauogle}) and
(\ref{taumacho}), is above the  expectations from 
all these models, it was suggested
that the cause of these large rates could be the fact that the bulge
is actually triaxial, with the larger axis making a small angle with
respect to the l.o.s. \cite{ki94,pa94}. In this way, the average
distance between sources and lenses is larger, with a corresponding
increase of the associated Einstein radii, and also the l.o.s. to a
source goes through a larger number of lenses. If this is the case, 
ML searches
would have rediscovered that our Galaxy is a barred spiral (see Section
2.5). In addition, the velocities of stars in a bar would be smaller
in the directions orthogonal to the major axis, helping to explain
the event durations observed, i.e. $T\sim 10$--50~d, which would be
too long for faint stars in a spherically symmetric \cite{gi94} or
axisymmetric \cite{ki94} bulge model. The predictions for barred
models were discussed in \cite{zh95,ha95,ev94c,ga95c,bl95,zh96}, 
and if the total bar mass
is  close to the maximum values consistent with the dynamical
estimates, $M_{bar}\simeq 2\times 10^{10}M_\odot$, and the angle between
the major axis of the bar and the l.o.s. is close to its minimum value
consistent with the Dwek et al.  inferred range, 
$\alpha=20^\circ\pm10^\circ$ \cite{dw95},
the lensing from bar objects, added to the maximum disk faint star
contribution, may account for the observed optical depth.
 If there is dark matter in the disk (thin or thick), this
can further help to increase $\tau$, as is also apparent from
Table~2. Furthermore, due to the small velocity dispersion of the disk
constituents, the event durations can be in the range of the observed
ones even for masses near the Hydrogen burning limit \cite{de95b}. 

If the lenses (and sources) belong to a bar, one expects to find
an asymmetry between the rate predictions for positive and negative
longitudes, and this signature may be searched for by mapping the rates in
different fields. One may also observe a magnitude offset between
the stars observed at opposite longitudes, with those at negative $\ell$,
which are further away, being fainter. This has actually
been observed by the OGLE team \cite{st94}. On the other hand, 
 sources in a bar are more likely to be lensed if they are located  on 
the far side, and hence 
there may also be an observable magnitude offset of the
lensed stars with respect to the whole sample in a given field
\cite{st95}. 
It has also been suggested that ML searches in the infrared, which
allow one to see in fields which are otherwise obscured, may help to better
distinguish disk and bulge lenses by observing the very inner region
of the bulge \cite{go95b}.

One complication in the interpretation of the bulge results is that a
non--negligible fraction of the source stars actually belong to the
disk \cite{mo96}. For instance, Terndrup \cite{te88} 
estimated that $\sim15$\% of
the giant stars in Baade's Window are in the disk, and this fraction
may be even larger for main sequence stars. Furthermore, in fields at
larger longitudes and similar latitudes, the number of disk stars is
almost unaffected while that of bulge stars decreases significantly,
making the effects of disk stars relatively more important in those
fields. The different distribution and motion of this second source
population affects the ML predictions. Their main implications are
that, due to the smaller  optical depth associated with the foreground 
disk sources, the value of the optical depth of bulge sources
is actually larger than the total average depth. If the disk is not
hollow, the contribution to the rate from disk sources is
non--negligible, and since events where both the lens and the source
are in the disk have particularly long durations, this can help to
account for some of the large $T$ events observed (which actually seem
to be concentrated towards small latitudes \cite{al96}). Also, since
the optical depth of disk sources lensed by objects in a bar  should
be larger at positive longitudes (unlike the bar--bar events), the
asymmetric signatures of the ML maps will be reduced.

There have been attempts to estimate the lens masses from the observed
event durations, and these do not hint towards a significant brown
dwarf contribution, but rather suggest lens masses  around
0.1--0.5~$M_\odot$ \cite{zh95,ha95b,de95b,zh96,al96}. However, one should keep in
mind that the mass function of the two lens populations, i.e. the 
disk and bulge stars, need not be similar, and this further
complicates the interpretation of the observations.

\section{Further determination of the lens parameters}

In Section 3.1 we have obtained the expected magnification of the 
source star luminosity during an ML event in the 
approximation that the source is point--like and that the lens has a 
uniform motion with respect to the ML tube. In this case, 
all of the information about the physical parameters of the lens is
contained in the event duration $T$, which depends on the mass $m$, the
distance to the observer $D_{ol}$, the relative
velocity in the plane orthogonal to the l.o.s., $v^\perp$, as
well as on the distance to the source, which is sometimes (e.g. in
bulge observations) not very well constrained.

Some methods have been proposed for obtaining additional information 
about the lens parameters.
 These include parallax measurements, which 
generally require the use of a satellite so as to have different observing 
positions, unless the event is long 
enough so that the Earth's motion induces a sizeable variation of the
relative lens velocity during the event. 
Another possibility arises when the effects of the finite 
size of the source are detectable, which allows to determine the 
lens proper motion. Finally, in the case that the source or the lens are 
binary objects, the resulting light--curve is much more complicated and
it is possible to learn more about the lens parameters.

\subsection{Parallax measurements}

The idea of performing simultaneous observations from ground and 
space--based telescopes of an ML event was proposed by Refsdal
\cite{re66} as a way of constraining the distances and masses of the lensing
objects. 
Grieger, Kayser and Refsdal \cite{gr86} proposed 
to measure the parallax effect 
on lensed quasars in order to gain information about the relative 
transverse velocities involved (see also \cite{ka86}). 
Gould proposed to apply the same ideas 
 to the ongoing searches towards the LMC and the bulge
\cite{go92b,go94b}. Parallax observations allow one to determine the
projection of $R_E$ onto the observer's plane, i.e. the so called
reduced Einstein radius
\begin{equation}
\tilde{R}_E\equiv R_E \frac{D_{os}}{D_{ls}}.
\end{equation}
Combining this with the information about the event duration, one can
obtain the modulus of the reduced velocity
\begin{equation}
\tilde{v} \equiv \frac{\tilde R_E}{T}=v^\perp \frac{D_{os}}{D_{ls}},
\end{equation}
which is a kinematical variable independent of the lens mass.

The strategy for measuring the parallax is the following: 
if the lensing event is monitored from two 
telescopes with no relative motion between them, the 
position of the lens with respect to the l.o.s. to the star 
from the Earth telescope in units of the Einstein radius can be 
expressed as 
\begin{equation}
{\bf u}(t)=\frac{{\bf b}}{R_E} +\frac{{\bf v}^\perp}{R_E} (t-t_0),
\label{u}
\end{equation}
and that from the satellite telescope as
\begin{equation}
{\bf u}'(t)=\frac{{\bf b}'}{R_E} +\frac{{\bf v}^\perp}{R_E} (t-t'_0),
\label{u'}
\end{equation}
with ${\bf b}\cdot{\bf v}^\perp = 0 ={\bf b}'\cdot{\bf v}^\perp$. From 
the observed light--curves $A(u(t))$ and $A(u(t'))$, the values of 
$t_0$, $t'_0$, $u_{min}=l/R_E$, $u'_{min}=l'/R_E$ and $T$ can be 
obtained. Denoting by ${\bf r}$ the projection orthogonal to the line 
of sight of the satellite position with respect to the Earth, 
${\bf u}(t)$ and ${\bf u}'(t)$ can be related through
\begin{equation}
{\bf u}'(t)={\bf u}(t) + \frac{{\bf r}}{\tilde R_E} .
\label{r}
\end{equation}
Combining Eqs.~(\ref{u}), (\ref{u'}) and (\ref{r}), we obtain for
$\Delta {\bf u}\equiv{\bf u}'-{\bf u}$ that
\begin{equation}
\Delta {\bf u}= \frac{{\bf r}}{\tilde R_E} =
{\bf u}'_{min} - {\bf u}_{min} - \frac{{\bf v}^\perp}{R_E}(t_0'-t_0).
\end{equation}
The fit to the light--curves provides a measurement of the modulus 
$u_{min}$, $u'_{min}$ and $v^\perp/R_E =T^{-1}$, but not of the 
directions. Thus, $\Delta u \equiv |\Delta {\bf u}|$ can only be 
determined up to a two--fold degeneracy
\begin{equation}
\Delta u_{\pm}= \sqrt{(u'_{min} \pm u_{min})^2 + \frac{(t'_0-t_0)^2}
{T^2}}.
\end{equation}
This corresponds to two possible values of the reduced Einstein radius
\begin{equation}
\tilde R_{E\pm}=\frac{r}{\Delta u_\pm},
\end{equation}
and of the modulus of the reduced velocity
\begin{equation}
\tilde{v}_{\pm} =\frac{r}{T \Delta u_{\pm}}.
\end{equation}

The resuting components of $\tilde{{\bf v}}_{\pm}$ in the direction 
parallel and perpendicular to ${\bf r}$ are
\begin{eqnarray}
\tilde{v}_{\pm}^\parallel&=&\frac{r (t'_0 -t_0)}{T^2(\Delta 
u_{\pm})^2},\nonumber\\
|\tilde{v}_{\pm}^\perp|&=&\frac{r (u'_{min}\pm u_{min})}{T 
(\Delta u_\pm)^2}.
\end{eqnarray}
Various strategies have been proposed for lifting the degeneracy.
Gould \cite{go94b} suggested performing observations from a second 
satellite, which would allow an unambiguous determination of $\tilde{v}$,
 since only
one of the values determined from the second satellite will coincide 
with one of the values determined from the first satellite 
measurements. If the positions of the two satellites with 
respect to the Earth, ${\bf r}$ and ${\bf r}'$, are not parallel, the sign of 
$\tilde{v}$ can also be determined. 
However, since the launching of a satellite into solar 
orbit is already a difficult and expensive enterprise, the necessity 
for having a second one is not a convenient solution. Another proposal 
\cite{go95} is to try to measure the difference in the time duration 
as measured from the satellite and from the Earth, $\delta T\equiv
T'-T$, arising from the slightly different actual velocities of the
satellite and the Earth. If 
$\delta T/T$ is measured with enough accuracy, then $\Delta u$ can be 
unambigously determined (see also \cite{bo95b,ga95d}). 
This requires good coverage and accurate 
following of the lensing events both from the ground and from space. 
Now that the `Early Warning' systems \cite{pr95,ud94d} 
have shown the ability of detecting the 
lensing events in real time, while the magnification of the source is 
starting, the strategy would be to monitor several 
million stars from the ground 
 and to follow with the satellite telescope those sources 
that have been singled out by the Early Warning system as possible 
ML candidates.

Obtaining  $\tilde{R}_E$ and $\tilde{v}$ for a large fraction 
of the 
observed ML events would allow a better characterization of 
the lensing population and, hopefully, distinguish among some of 
the proposed scenarios. For example, for observations towards the LMC, it 
would be possible to distinguish between lenses in the Galaxy and lenses 
in the LMC. The measurement of very large values of $\tilde{v} = 
v^\perp/(1-x)$ would be evidence of lenses belonging to the LMC, for 
which it is expected that $\tilde{v}  \sim$ 1000 km/s, while 
significantly smaller values would result from Galactic lenses, for
which  
$\tilde{v}  \sim v^\perp$. Furthermore, in this last case, parallax 
measurements would provide direct information on the velocity 
distribution of the lenses,
 helping to distinguish between disk, spheroid and halo 
lenses.

For observations towards the Galactic bulge, the reduced velocity 
measurements would also be very helpful for distinguishing between
disk and  
bulge lenses \cite{ha95}. Typical disk lenses are located half--way 
to the bulge, thus the reduced velocity is approximately twice the 
velocity of the lens with respect to the ML tube. 
Furthermore, the flat rotation curve of the disk and the relatively 
small velocity dispersion of disk stars  gives an 
approximate relation between $v^\perp $ and the lens distance
(neglecting the motion of the source), 
$v^\perp  \sim v_{rot} D_{ol}/D_{os}$, where for bulge star 
sources $D_{os} \simeq R_0$. Thus, for a population of disk 
lenses, the reduced velocities are expected to be $\tilde{v}\simeq
v_{rot}D_{ol}/D_{ls}$, and hence to be close to 
200 km/s. On the other hand, a large enhacement factor for $\tilde{v} 
$ with respect to $v^\perp$ is expected for bulge lenses, as these 
would be located close to the source stars. Velocity dispersions of 
bulge stars of the order of 100 km/s give rise to reduced velocities
peaked at $\tilde{v}  \sim 1000$ km/s. Thus, for a large fraction 
of the events, it would be possible to determine whether they are due to 
disk or bulge lenses, and for events due to disk lenses, to obtain the most 
probable lens distance, which in turn helps to better constrain the 
lens masses. If the time duration and reduced velocity are known for a 
given event, the mass and distance of the lens are related through
\begin{equation}
m=\frac{\tilde{v}^2 T^2 c^2 D_{ls}}{4 G D_{os} D_{ol}}.
\end{equation}

\begin{figure}
\centerline{\psfig{figure=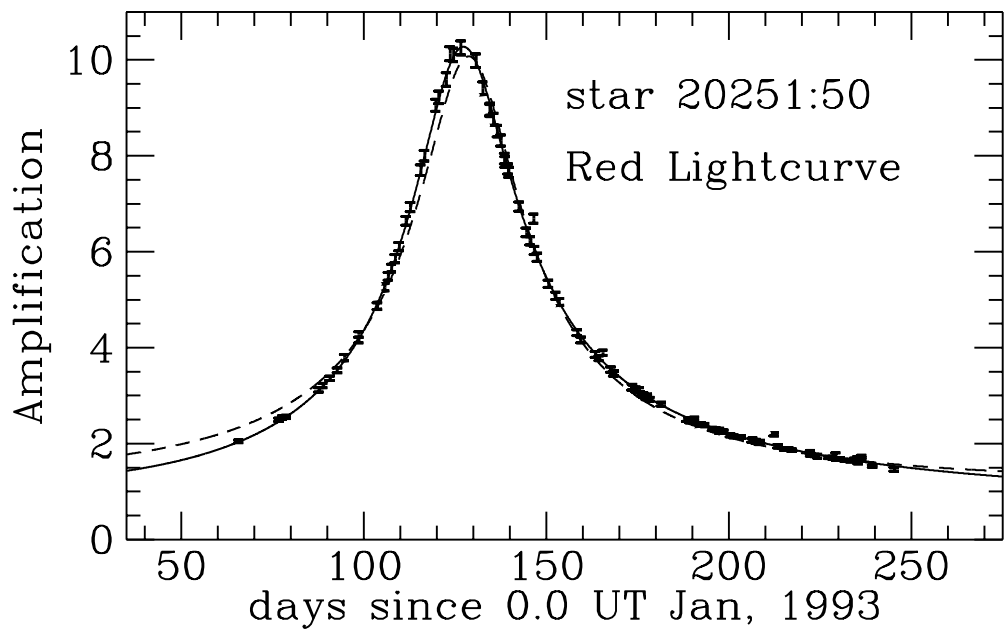,height=10cm}}
\caption{First observation of parallax in an ML event by the MACHO
collaboration.  The
dashed line is the fit without taking into account the Earth motion
while the solid line does include it.}
\label{parallax.fig}
\end{figure}

Parallax effects can be measured from the Earth only for very long 
duration events, for which the motion of the Earth gives rise to a 
significant correction to the  light--curve. An event of this kind has 
already been detected by the MACHO collaboration in their  
 first season of bulge observations \cite{al95d}. 
Their longest event, lasting 
110~days, presents a significantly asymmetric, but  
still achromatic, light--curve, as is expected to be the case when
the approximation of constant 
velocities of the lens, source and observer  does not hold. 
The correction for the orbital motion of the Earth significantly 
improves the fit to the observed light--curve (see
Fig.~\ref{parallax.fig}, taken from Ref.~\cite{be95b}). 
 The reduced velocity inferred is $\tilde
v=75\pm 5$~km/s, and it should be pointing at an angle of
28$^\circ$ with respect to the direction of increasing longitudes, supporting
the interpretation that the lens is in the disk. The best 
fitting values obtained for the lens parameters 
are $D_{ol}= 1.7^{+1.1}_{-0.7}$ kpc and $m = 
1.3^{+1.3}_{-0.6} M_{\odot}$, indicating that the lens is probably an old 
white dwarf or a neutron star in the disk\footnote{This fit neglects
the possible rotational motion of the source. 
If the source were moving in the direction
of the Sun's rotation, e.g. if it were a disk star or rotating bulge
star, the inferred value of $D_{ol}$ would be larger, and hence
the lens mass smaller \cite{mo96}.}. 
Hence, the observation of 
parallax in one event allows one to achieve a better knowledge of 
the lens characteristics than in 
the rest of the events. However, 
parallax measurements can be performed from the ground for only a small 
fraction of the events, those with very long durations.
For the majority of the events, which have  $T$ between 10 and 50
days, $\tilde{v} 
$ cannot be obtained from the ground and a space--based telescope 
is necessary.

\subsection{Proper motions}

For some particular ML events it may become possible to obtain
information on the Einstein angle $\theta_E$ defined by
\begin{equation}
\theta_E\equiv\frac{R_E}{D_{ol}}
\end{equation}
and hence know the `proper motion' 
\begin{equation}
\mu\equiv\frac{\theta_E}{T}=\frac{v^\perp}{D_{ol}},
\end{equation}
which is just the angular velocity of the lens relative to the
source. Notice that the proper motion is a kinematical variable,
independent of the lens mass, while $\theta_E$ is independent of the
lens velocity, and hence these parameters 
are less convoluted than the event duration
$T$. 

The clearest situation in which the proper motion can be measured 
 is when the radial 
extension of the source $R_s$  projected onto the lens  plane 
is comparable to the impact parameter of
the lens trajectory $b$, or equivalently that the angle subtended by
$R_s$, $\theta_s\equiv R_s/D_{os}$, is larger than the angle subtended
by $b$,
$\beta\equiv b/D_{ol}$ (see Fig.~\ref{angles.fig}). In this case, 
 the point--like source approximation breaks down, and 
 the light--curve of the ML
event is modified allowing the proper motion of the lens to be extracted
 from it \cite{ma73,go94,wi94,ne94}. One proper motion 
has been recently measured by
the MACHO collaboration in a bulge star \cite{pr95b}.

The magnification of a circular source with uniform surface brightness
lensed by a point--like mass has been obtained by Witt and Mao
\cite{wi94} (see also \cite{ma73}). If 
$\beta\gg \theta_s$, the
light--curve 
is almost identical to that of a  point--like source, but if $\beta$ is 
comparable with or smaller than $\theta_s$, the light--curve is
significantly different. When $\beta>\theta_s/2$, the peak 
magnification is larger than that for the point--like source, while  when 
$\beta\ll \theta_s/2$ 
the peak amplification is always smaller than in the point--like case. 
In particular, when the alignment is perfect ($\beta=0$), 
the amplification does
not diverge as the point--like source expression Eq.~(\ref{amp.eq})
indicates, but instead reaches a maximum value (assuming a uniform
surface brightness) given by
\begin{equation}
A_{max}=\sqrt{1+\frac{16 G m D_{ls} D_{os}}{R_{s}^2 c^2 D_{ol}} }.
\label{amax}
\end{equation}

Comparing the peak 
magnification with that expected from the point--like source light--curve 
determined by a fit to the measured amplifications  far from the peak, it 
is possible to have a measure of the ratio of the impact parameter to 
the projected source radius, 
\begin{equation}
Z\equiv b D_{os}/R_s
D_{ol}=\beta/\theta_s.
\end{equation}
 The point--like 
source peak amplification gives us the value of $u_{min} = b/R_E$ 
through Eq.~(\ref{amp.eq}). Now it is possible to determine the angle 
\begin{equation}
\theta_E= \frac{b}{u_{min}D_{ol}} = \frac{Z 
R_s}{u_{min}D_{os}},
\end{equation}
 as the source  radius $R_s$ and the distance $D_{os}$ can usually be 
estimated independently. Hence, the
`proper motion' is
\begin{equation}
\mu=\frac{\theta_E}{T}= \frac{Z 
R_s}{T u_{min}D_{os}}.
\end{equation}
As this effect appears when the lens transits the surface of the 
source, it is only expected to be present in some high magnification 
events, for which a better knowledge of the lens parameters would thus 
be possible. 
Microlensing searches focused on picking up this type of
events may then be particularly interesting \cite{go95h}. If all
events with $u_{min}<1$ (i.e. those with
$\beta<\theta_E$) can be detected, 
since $\mu$ can be measured only for those with
$\beta<\theta_s$, one has that the fraction of events with
measurable proper motion is $\simeq \theta_s/\theta_E\propto
1/\sqrt{m}$. For LMC observations, and assuming $m=0.1M_\odot$, this
fraction is $\simeq 10^{-3}$ for Galactic lenses and $10^{-2}$ for LMC
ones. The chances of measuring the proper motion are much better when
the lens is a binary system (Section~5.4), since during a caustic
crossing the light--curve is very likely to be affected by the finite
source size.

The measurement of the proper motion for LMC events would allow one to 
distinguish between LMC and Galactic lenses. LMC lenses would give rise 
to proper motions $\mu \leq 2$ km/s/kpc. Larger values 
would indicate a Galactic lensing population and could also help to 
distinguish among the different scenarios proposed. For bulge 
observations, somewhat shorter proper motions are expected for bulge 
lenses than for disk ones, but a clear distinction between the two
populations is not possible.

Some modifications of this method for measuring the lens proper motion 
have been proposed. Witt \cite{wi95b} suggested looking for color 
changes in the light--curve arising from  the fact that the 
brightness profile of an extended source 
depends on the wavelength. Gould and Welch \cite{go95g} estimated that 
photometry in two 
bands (optical and infrared) would allow one to measure proper motions for 
lensing events with impact parameters up to two projected source 
radii, doubling the number of possible proper motion measurements. 
It has also been pointed out by Loeb and Sasselov 
\cite{lo95} that narrowband photometry of a giant source lensing event 
can be used to measure the proper motion  and the angular radius of the 
Einstein ring of the lens. 
Large radius giants have narrowband images 
with a thin ring shape and their lensing gives rise to a light--curve 
with two sharp peaks which can be used to measure the proper motion.
A related proposal, by Simmons, Willis and Newsam \cite{si95}, is to 
measure the time dependence of the polarization of the  light from the
source 
which results from the fact that the light from different parts of the 
limb (which is sometimes polarized) will be magnified by different 
amounts during a lensing event, providing a measure of the 
proper motion. Maoz and Gould 
\cite{ma94} proposed exploiting the fact that, during the lensing of 
an extended rotating source, the differential amplification of the 
source disk induces a shift in the spectral lines. Spectroscopic 
observations whith a large telescope can detect the time shift of the 
spectral lines and the proper motion of the lens can then be determined. 
 Yet another proposal by Han, Narayan and Gould is to
measure the diffraction pattern of a bulge ML event suffering occultation by
the Moon, which may be sensitive to the separation $\Delta\theta$ in 
Eq.~(\ref{dtheta.eq}), and this could allow one to infer the angle
$\theta_E$. 

Another method proposed for measuring the angular
Einstein ring radius \cite{wa95} is
based on the fact that, although the two ML
images cannot be resolved, the apparent motion of the centroid of
images may be detected. The centroid trajectory follows an ellipse
whose size is proportional to the Einstein angular radius $\theta_E$,
and for large amplification events the maximum centroid deflection is
$\theta_E/\sqrt{8}$. In order to observe this small effect, 
 a satellite experiment
with a 1~m telescope and  very high accuracy in the photometry would
be required.
Finally, using the next generation of optical interferometers, it may
become feasible to measure by direct imaging the Einstein angles of
some of the long duration events (corresponding to
massive lenses) towards the bulge.

Let us note that if, in addition to the proper motion, also
the reduced velocity is measured for the same event using parallax, 
then the three 
parameters $D_{ol}$, $m$ and $v^\perp$ can be determined.

\subsection{Binary sources}

Up to now, we have only considered the case in which both the source 
and the lens are single objects. However, it is well--known that most 
stars are members of binary systems. Abt estimated that the 
fraction of binary stars is between $60\%$ and $100\%$ \cite{ab83}. 
Thus, it is important to take into account the modifications that 
could appear when the source or the lens are binary systems, 
as some of the discussed signatures of an ML event may be 
lost in these cases. 

The case of binary source stars has been analysed in detail by Griest 
and Hu \cite{gr92}. The light--curve can in this case be quite 
different from the one for a single source. If both stars are lensed, the 
light--curve can present two peaks or be quite time asymmetric. When 
only one of the stars is significantly lensed, the light--curve will 
still be symmetric, but will 
have an offset due to the luminosity of the companion if the two stars
 cannot 
be resolved. If the dimmer star is the lensed one, the shape of the 
light--curve may be significantly affected. On the other hand, if the 
two stars have different colors and only one is lensed, then the 
apparent color of the unresolved system will change in time and the 
achromaticity signature may be lost. The total ML rate is 
expected to be a factor $5\%$ to $10\%$ larger for binary sources.

For two point--like sources of apparent luminosities $L_1$ and $L_2$, the 
individual amplification of each one has the usual expression in terms 
of the distance of the lens to the l.o.s. scaled to the 
Einstein radius, $A_1(t)=A(u_1(t))$ and $A_2(t)=A(u_2(t))$. The total 
amplification is given by
\begin{equation}
A=\frac{A_1 L_1 + A_2 L_2}{L_1 + L_2}= A_1 (1-s) + A_2 s,
\end{equation}
with
\begin{equation}
s\equiv \frac{L_2}{L_1+L_2}.
\end{equation}
From this expression one can  obtain the magnification as a 
function of time for the different possible geometries involved. 
Griest and Hu \cite{gr92} described a huge variety of possible 
light--curves,  estimating that for a fraction between $10\%$ 
and $20\%$ of the LMC events, the light--curves may probably be 
distinguishable from the single source ones, and hence these events 
could be rejected as ML candidates in a microlensing search 
requiring a good fit to the single source light--curve. For most of 
the events, however, the difference will be small. The fraction of
genuinely binary events depends on the stellar type considered and,
for example, for giant sources it is not expected that a companion star
would be bright enough to significantly affect the light--curve. Hence, for 
bulge observations, the fraction will be smaller.

When the two stars have different colors, the factor $s$ depends on 
the color. Thus, the total amplification will also depend on the 
color, and the ML event will not look achromatic. However, 
Griest and Hu \cite{gr92} estimated that the achromaticities are
 expected to be 
small in most of the cases. They also noted that if the color shift 
is measured as a function of time, combining this with the light--curve 
information would allow one to determine the trajectory of the 
lens projected into the source plane, up to a twofold degeneracy 
arising from 
the reflection symmetry of the geometry. This could also be done for a 
double--peaked light--curve (in which both stars are subsequently
magnified),  even without a color shift measurement.
The reduced velocity of the lens could also be determined in these 
cases.

\subsection{Binary lenses}

The possibility that the lens consists of a binary system has also been 
widely discussed in the literature \cite{sc86b,wi90,go92,wi95,ma95}. 
The resulting light--curves may have a much richer structure in this 
case.

In order to extend the formalism presented in Section 3.1 to the case of 
two point--like lenses of masses $m_1$ and $m_2$, it is useful to work
in the complex notation, with adimensional variables,
 proposed by Witt \cite{wi90}. We denote $z = x
+{\rm i}y$, where $(x,y)$ are Cartesian coordinates in the deflector
plane, and $\zeta = \xi +{\rm i}\eta$, with $(\xi,\eta)$  coordinates 
in the source
plane. The variable $z$ is normalized to the Einstein
radius $R_E$ corresponding to the total mass $m_T=m_1 +m_2$ and $\zeta$ to its
projection onto the source plane. We denote by $M_1$ and $M_2$
the masses of the deflectors normalized to the total mass,
$M_i=m_i/m_T$, and $z_1$ and $z_2$  their positions. 

A light ray crossing the lens plane at a point $z$ will be 
deflected by a (complex) angle corresponding to the superposition of 
the deflections produced by the two individual masses
\begin{equation}
\alpha =\frac{D_{os}}{D_{ls}} \left(\frac{M_1}{\bar{z}-\bar{z}_1}+
\frac{M_2}{\bar{z}-\bar{z}_2}\right),
\end{equation}
where the angle $\alpha$ is here normalised  to the angular Einstein radius 
$R_E/D_{ol}$, and we have used Eq.~(\ref{alpha.eq}) written in terms
of the new variables just introduced.

For a source located at $\zeta$, the image positions in the lens plane 
$z_i$ satisfy \cite{wi95}
\begin{eqnarray}
\zeta&=&z_i-\alpha \frac{D_{ls}}{D_{os}}\nonumber\\
&=& z_i+\frac{M_1}{\bar{z}_1-\bar{z_i}}+\frac{M_2}{\bar{z}_2-\bar{z_i}}.
\label{zeta}
\end{eqnarray}
In order to obtain the image positions $z_i$, it is necessary to 
invert Eq.~(\ref{zeta}). As this is an equation of fifth degree in
$z_i$,  
it is not possible to do it analytically for arbitrary values of 
$\zeta$. This constitutes the major obstacle to the analytical 
treatment of the binary lens case.

The magnification of the image $z_i$ is given by
\begin{equation}
\left. A_i=\frac{1}{|{\rm det} J|}\right|_{z=z_i},
\end{equation}
where
\begin{equation}
{\rm det} J= 1-\frac{\partial \zeta}{\partial \bar{z}} 
\overline{\frac{\partial \zeta}{\partial\bar{z}}}
\end{equation}
is the determinant of the Jacobian. For certain values of the source 
and image positions, the determinant vanishes and the amplification 
factor diverges. These source and image positions with diverging
amplifications form closed curves 
and are called caustics and critical curves respectively. When a 
source crosses a caustic, a pair of images appears or disappears. It 
has been shown \cite{sc86b} that the lens equation (\ref{zeta}) has either 
three or five solutions depending on whether the source is outside or 
inside the caustic.

Even if the image positions cannot be found analytically, Witt and 
Mao \cite{wi95} have developed a method for obtaining the magnification of 
the images without knowing their positions. They also showed that the 
minimum amplification for a source inside the caustic is $A=3$. A 
minimum magnification below 3 between two caustic crossings indicates 
that either there is an extra unlensed source which is not 
resolved, or that the lensing agent consists of three or more stars.
However, the later possibility is extremely unlikely because such a
configuration would be gravitationally unstable.

Mao and Paczynski \cite{ma91} estimated that  approximately $10\%$ 
of the events observed towards the Galactic bulge should show strong 
effects in their light--curves due to the lens being a binary. 
These light--curves can be quite 
asymmetric and present multiple peaks. The magnification is expected 
to be achromatic. A binary lens event can be characterized by the 
total mass of the system $m_T$, the  ratio of the 
individual masses $q=m_1/m_2$, their separation projected onto the 
lens plane $a$ and the parameters describing the trajectory: the 
impact parameter $b$, the transverse relative velocity $v^\perp$ 
of the center of mass 
 and the angle  $\theta$ between 
the projected axis of the binary and the transverse velocity vector.
The variety of light--curves which can result 
from ML by a binary system is then very large.
 Mao and Di Stefano 
\cite{ma95} developed a best fitting method of the observed
light--curve  to the binary lens theoretical curve which allows one 
to determine 
the parameters  $u_{min}\equiv b/R_E$, $t_0$, $T\equiv R_E/v^\perp$,
as in the single lens case, 
plus the three additional  parameters $a/R_E,$ $q$ and $\theta$.
 When performing these fits, it is necessary also to 
allow for a fraction of the light contribution to come from unlensed 
sources which are not resolved. The time duration $T$ fixes  
the relation between the 
total mass $m_T$, $v^\perp$ and $x$, as in the single lens case. 
Additional information about them can be obtained only 
in the case in which the detailed behaviour of the magnification of a 
peak is well sampled so as to allow to resolve the source structure
during the caustic crossing, making then possible the determination of  
 the proper motion.

As the magnitude spikes can have short durations, it is necessary to 
have a good coverage of the event in order to detect the features in 
the light--curve. This is now easier with the Early Warning systems 
already at work and the organised followup networks.  
Two ML events by a binary lens have already 
been detected towards the bulge, the first by the OGLE group 
\cite{ud94c}, and confirmed by MACHO \cite{be95b}, and the second 
one by the DUO project \cite{al95a}. 

\begin{figure}
\centerline{\psfig{figure=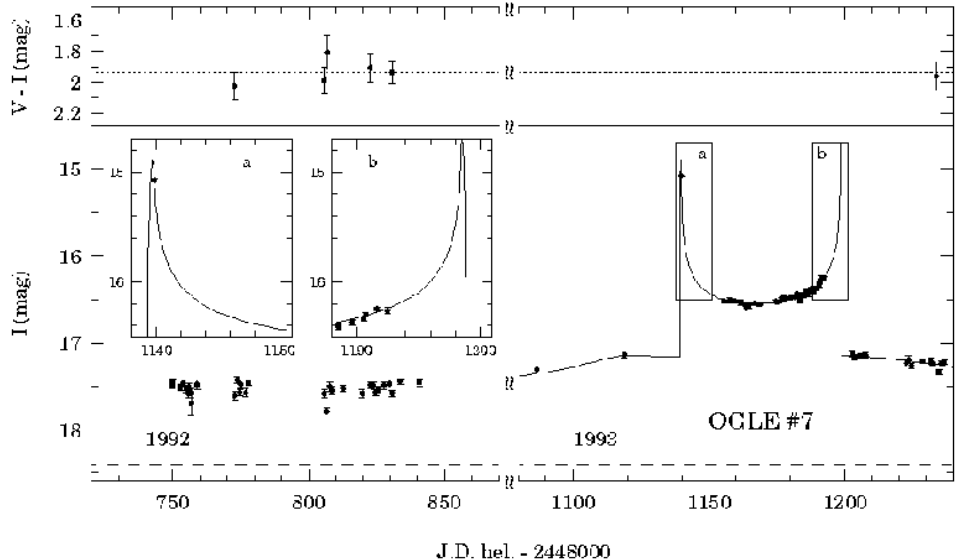,angle=-90,height=10cm}}
\caption{ML event due to a binary lens observed by the OGLE
collaboration, reprinted with permission from The
Astrophysical Journal.}
\label{ogle7.fig}
\end{figure}

The first binary lens event 
detected by OGLE is shown in Fig.~\ref{ogle7.fig} (taken
from Ref.~\cite{ud94c}). It 
 presents two maxima and a characteristic  U--shaped plateau 
between them which is due to
the extra pair of images produced when the source
crosses  the caustic. The best fitting model consists of 
two nearly equal mass lenses ($q=1.02$), separated by a distance of 
the order of the total mass Einstein radius ($a/R_E=1.14$), with the 
center of mass passing close to the l.o.s. to the source 
($b/R_E=0.05$) at an angle $\theta=48^\circ.3$. It was a long duration 
event with $T=80$~d. The estimated contribution by the lensed star 
to the light--curve at the baseline is $f=56\%$. The MACHO data 
\cite{be95b} provides a nice 
confirmation of the event with a very good coverage of the second 
caustic crossing which was largely missed by OGLE. Fitting the second 
caustic crossing points to the magnification near a caustic for a 
finite size source star with a  realistic limb darkening  
 \cite{sc87}, they fixed the time that the diameter of the 
star took to cross the caustic to be 10 hours, and from this proper
motion measurement they estimated the lens
relative velocity to be $v^\perp = x (48$ km/sec)$ R_{s}/R_\odot$. 

The second event, found by DUO, 
presents three different peaks with a plateau between 
the first two \cite{al95a}. In the best fit model, the source 
trajectory has two caustic crossing and passes near a cusp, where the 
third peak arises. The time duration is $T=8.5$ days, suggesting that
the total mass is $\sim 10^{-2}M_\odot$. The ratio of masses is
$q\simeq 0.33$, so that the mass of the 
lighter lens is close to that of Jupiter. The 
 size of the source is not well constrained  in this case, due to poor
sampling near the caustic crossing. The estimated fraction of 
the light contribution at the baseline from the lensed source is 
$f=70\%$. Further analysis of the data showed some evidence of a shift 
in the light centroid, as expected  when 
blending with an unlensed source is present. This hypothesis and the 
estimated location of the unlensed star are confirmed by two good 
quality images of the object, obtained a posteriori, in which the two 
stars are resolved. 

Another situation in which binarity could play a role is the 
possibility of ML events in which one member of a binary is 
lensed by its companion \cite{ma73,go95c}. 
Gould \cite{go95c} estimated that the optical 
depth of these processes is extremely small, $\tau \leq 10^{-11}$, for 
binaries in which at least one member is an ordinary (non--compact) 
star.

A further possibility which  has been analysed is that planetary systems 
of Galactic disk stars might be detected by ML of stars in the 
Galactic bulge \cite{ma91,go92}.  Typical 
planetary signals produce a deviation from the standard, single lens, 
light--curve for only a small fraction of the event duration time. The
observation of these features would allow one to determine the ratio 
of the planet and star masses.
Gould and Loeb \cite{go92} estimated 
that for about $20\%$ of the lensing events in the bulge due to
Solar--type disk stars having Jupiter--like planets, there  
should be a noticeable 
signature ($>5$\% amplification) of planetary lensing in their light--curves. 
Due to the short duration of the planetary 
signal (1~d for a Jupiter mass planet),
 several observations per day are required  in order to 
pick it up. The GMAN (Global Microlensing Alert Network) \cite{pr95}
and PLANET (Probing Lensing Anomalies Network) collaborations already
started the intense scrutiny of alerted events to look for such
signals, using telescope networks around the globe.

Hardy and Walker \cite{ha95e} have proposed  looking for parallax
effects in binary lens events using three 1~m ground based
telescopes. The fact that the light--curve is very steep when the limb
of the source touches the caustic allows a quite precise determination
of the location of the source with respect to the caustic. This would
allow one 
to observe parallax effects by combining observations 
from distant Earth locations. In this way
the angular Einstein ring radius divided by the total lens mass,
$\theta_E/m_T$,  can be determined. Since a detailed
measurement of the caustic crossing may also allow one to infer the
proper motion, these observations could determine the three lens
parameters, $m$, $v^\perp$ and $D_{ol}$ simultaneously.

Finally, another method for extracting 
information about the lens parameters for an ordinary ML event, in the 
case in which both the lens and the source are main--sequence stars, has 
been proposed by Kamionkowski \cite{ka95,bu96}. If the light from the lens 
contributes significantly to the observed brightness and it has a 
different color from the source light, it is expected that for some of the 
events  a shift in the color can be measured. In this case, it would 
be possible to infer the lens and source colors and luminosities, and
hence the lens mass and the distances to the source and 
the lens if they are main--sequence stars. Using then the event
duration one may determine  the lens
velocity. 
 However, it would be difficult to be sure that the 
blending in the light is due to light from the lens rather than
 from another unresolved source. A discussion of the effects of
blending in ML searches can be found in Ref.~\cite{di95}.

\section{Additional searches}
In this Section we discuss the method of searching for ML of unresolved
stars which is being applied to ML searches in the Andromeda
galaxy. We also discuss the constraints on the number of red and brown
dwarfs in the Galaxy, which can be obtained with IR searches and which
give important information, complementary to that obtained with ML
searches, about the possible Galactic lensing populations.

\subsection{Microlensing of unresolved stars}

The standard microlensing technique discussed so far consists of 
monitoring the light flux from a large number of stars, looking for 
magnifications consistent with the ML event characteritics. This 
is the method used by the ongoing searches towards the Magellanic 
Clouds (MACHO and EROS) and the Galactic bulge (OGLE, MACHO and DUO). 
While this kind of search has led to spectacular results, it has 
the restriction that it can only be applied to targets  where 
several million stars can be resolved.

A different approach has been proposed by 
Baillon et al. \cite{ba92b,ba93} and by Crotts \cite{cr92}, 
 which consists in the search
 for ML of unresolved 
stars. The idea is to study the flux of light received by every CCD 
pixel in the picture of a galaxy, rather than the flux of the 
individual stars. This allows one to detect the amplification due 
to the lensing of one of the many stars present in a crowded field, even if 
the star is not resolved. The number of stars effectively monitored is 
greatly enlarged with this method. However, ML events are 
harder to detect since the light of each star contributes only a small 
fraction of the light in a pixel. Thus, only lensing of bright stars 
or very small impact parameter (high amplification) events in the
lensing of fainter stars can be detected. 
It has been proposed that this method,  applied to LMC searches, 
would give an order of magnitude more events than the standard 
technique \cite{bo93}. However, the main advantage of the pixel 
lensing method is that it allows one to look for ML events 
in more distant galaxies, where the standard lensing technique is not 
applicable due to the crowding limit. An interesting target for this 
kind of search is the Andromeda galaxy (M31) \cite{cr92,ba93}, which is 
located at a distance of $\sim 770$~kpc from us. These observations are 
not only sensitive to ML by compact objects in the Milky Way, 
but also to those in Andromeda itself. Observations towards M87 in the
Virgo cluster may be sensitive to intra--cluster dark matter also
\cite{go95j}. 

The theory and observational prospects for pixel lensing searches have 
been studied in Refs.~\cite{ba93,co95b,an95c,go95i,ha95d}; we present here a 
brief discussion. The basic idea of the method is that if a
star of flux $F_\star$ is lensed in a crowded field, by subtracting
from the actual images of the field a reference one (obtained by  
averaging several images taken before or after the event), one should
get just a stellar image with flux equal to $(A-1)F_\star$ on top of
the photon noise. The extent of the star image is fixed by the point
spread function (PSF), determined in turn by the actual seeing
conditions. If we denote by $\Omega_{PSF}$ the effective solid angle
associated with the spread in the star image, and by 
$\Sigma=\Sigma_{sky}+\Sigma_{gal}$ the background flux per unit solid
angle 
coming from the sky and from the galaxy light in the region where the
ML event is taking place, the signal to noise of the ML observation in
an exposure time $t_{exp}$ is just
\begin{equation}
\frac{S}{N}=\frac{(A-1)F_\star
t_{exp}}{\sqrt{\Sigma\Omega_{PSF}t_{exp}}}.
\label{sdn.eq}
\end{equation}
For a required value of $S/N$, there is thus a minimum amplification
$A_{min}$, which depends on $F_\star$, for the event to be detected.

The strategy adopted, for instance, by the AGAPE group is to study the
light--curve of superpixels, i.e. groups of pixels chosen so as to
approximately match the PSF of the stellar images, and look for
consecutive increments of the signal significantly above the noise to
identify a stellar variability. One may also trigger on the signal to
noise of the whole event, as advocated in \cite{go95i,ha95d}.

There are, of course, sources of fluctuations which have to be 
controlled for a successful observation. These include the limited 
 precision in the telescope pointing (which requires a geometric
realignment of the images), the changes in the background sky light
and in the transparency of the atmosphere (requiring a photometric
realignment of the images) and finally, to eventually convolve the
reference image with the actual seeing conditions.
 Some techniques have been developed to correct for these effects 
 \cite{an95c,to96}, and it is expected that they  can be kept smaller 
than the statistical fluctuations in the number of photons.

As for the standard ML searches, the main background for  
lensing events of unresolved stars 
is given by variable stars. The criteria to 
distinguish them are also the symmetric shape of the light--curve and 
the non repetition of the ML event. Also the achromaticity 
signature can be used \cite{an95c}: even if the lensing of a star will 
cause a color variation in the pixels involved, the ratio of the 
increment in the flux  from the reference flux in two different 
colors
\begin{equation}
\frac{(F-\langle F_{ref} \rangle)_{red}}
{(F-\langle F_{ref} \rangle)_{blue}}=
\frac{(F_{\star})_{red}}{(F_{\star})_{blue}}
\end{equation}
is constant in time during an ML event.

Another point which has to be taken into account in ML 
searches in distant galaxies is that the effect of the finite size of 
the source gives rise to a maximum amplification,  given by
Eq.~(\ref{amax}), as discussed in the  previous section.
Thus, for sources of a given radius $R_{s}$, 
there is a minimum lens mass giving rise to an observable event
\begin{equation}
m_{min}= 0.7 \times 10^{-7} M_\odot (1+A^2_{min}) 
\left(\frac{R_{s}}{R_\odot}\right)^2 \left(\frac{1\ {\rm kpc}}{D_{ls}}
\right),
\end{equation}
where $A_{min}$ is the threshold amplification associated to a given
$S/N$ requirement.
 This has the effect of reducing the number of detectable  
lensing events of red giant sources and the efficiency for lensing by
 very light objects in M31 \cite{ba93}.

Using Monte Carlo simulations and semi--analytic computations, Baillon et 
al. \cite{ba93} estimated that in searches towards M31 the best 
sensitivity is achieved for lens masses in the range 
$10^{-3}$--$10^{-5} M_{\odot}$, 
that most of the detectable events correspond to 
mild amplifications of bright stars rather than high amplifications of 
faint ones, and that feasible observation programs would lead to a few 
tens of events per year for brown dwarf masses in the range $10^{-5}$ 
to $10^{-1} M_\odot$ making up the halos of the Milky Way and M31. 
Similar conclusions were reached in
Ref.~\cite{co95b}. 

Two groups have already started observational programs towards M31
with the aim of proving the feasibility of the pixel lensing technique
and have already been able to identify variability of unresolved stars. 
AGAPE (Andromeda Galaxy Pixel Experiment) is a French collaboration 
that has  taken data using a 2~m telescope with a CCD 
camera at the Pic du Midi observatory during observing periods in 
1994 and 1995 \cite{an95c}. The Columbia--VATT collaboration 
is an experiment led by Crotts using mainly the Vatican Telescope in
Arizona \cite{to96}. 

\subsection{Microlensing towards Andromeda}

The Andromeda galaxy provides a very interesting target for 
microlensing. It is the closest large galaxy, and 
as it is located forming an angle of $119^{\circ}$ with respect
to the Galactic center, 
it provides a possibility for testing the dark constituency of the 
Galaxy in a direction complementary to the Galactic bulge and LMC 
targets. Moreover, the dark constituency of M31 itself can be probed,
and the fact that we see the disk of M31 with a high 
inclination ($i \simeq 75^{\circ}$ respect to the face on position) 
gives rise, for lenses in M31,  to a much larger 
optical depth for ML of sources in the far side of the 
disk than for those in  the near side \cite{cr92,ba93,gi94}.
This effect arises 
because the l.o.s. to the far side of M31 goes through much of the 
inner and denser part of the M31 halo and spheroid, and provides an 
interesting signature of ML. 

The probability of observing an ML event in a crowded field can 
be estimated as follows. First, we estimate the probability that a 
star of a given luminosity $L$ in the field gives rise to an 
observable event, and then sum over all the stars
\begin{equation}
\tau_{pixel}=\sum_{stars}\tau(L).
\end{equation}
Here  the sum over the stars is performed by integrating over the 
luminosity function (normalized to the surface brightness of the 
galaxy). This was performed for the case of M31 in
Refs.~\cite{ba93,co95b} using the available measurements for red
giants and a matching for fainter stars with the luminosity function
in the solar neighbourhood conveniently normalised.
To observe an ML event of a star of 
luminosity $L$, the mean amplification during the exposure has to be 
larger than a threshold amplification $A_{min}$ given by the minimum $S/N$
required through Eq.~(\ref{sdn.eq}), or 
equivalently the lensing object has to be at a distance from the l.o.s. 
smaller than the corresponding $u_{max}(L)$ obtained from
Eq.~(\ref{amp.eq}).  Thus, the probability that a star of luminosity $L$ be 
lensed to observable levels 
is just the number of lenses inside a tube of radius 
$u_{max}(L) R_E$ around the l.o.s., that is $u^2_{max}(L) 
\tau_{\star}$, where $\tau_{\star}$ is the optical depth for the 
standard ML of a star as given by Eq.~(\ref{opt.eq}).

Thus, the optical depth for unresolved stars can be estimated as
\begin{equation}
\tau_{pixel} = \tau_{\star} \sum_{stars} 
u^2_{max}(L).
\end{equation}
If the lenses belong to M31, the 
dependence of $\tau_{\star}$ on the observation position in M31 can 
be readily obtained from Eq.~(\ref{opt.eq}), using for $\rho(r)$ the density 
of M31 lensing objects, with $r$ being the distance of the lens to the 
M31 center, 
\begin{equation}
r=\sqrt{(D_{os}-D_{ol}-d \cos \Phi \tan i)^2+d^2},
\end{equation}
where $i \simeq 75^{\circ}$ is the inclination angle of the M31 disk, 
$d$ is the distance between the l.o.s. to the source and the M31 center, 
and $\Phi$ its angle to the far minor axis. Crotts \cite{cr92} 
obtained the resulting $\tau_{\star}$ for a spherically symmetric 
halo. The non-spherical axisymmetric case has been analysed in Ref. 
\cite{je94}. The contribution from a heavy
spheroid has been computed in Ref.~\cite{gi94}, while that of a disk 
population has been obtained in Ref.~\cite{go94d}.

\begin{figure}
\centerline{\psfig{figure=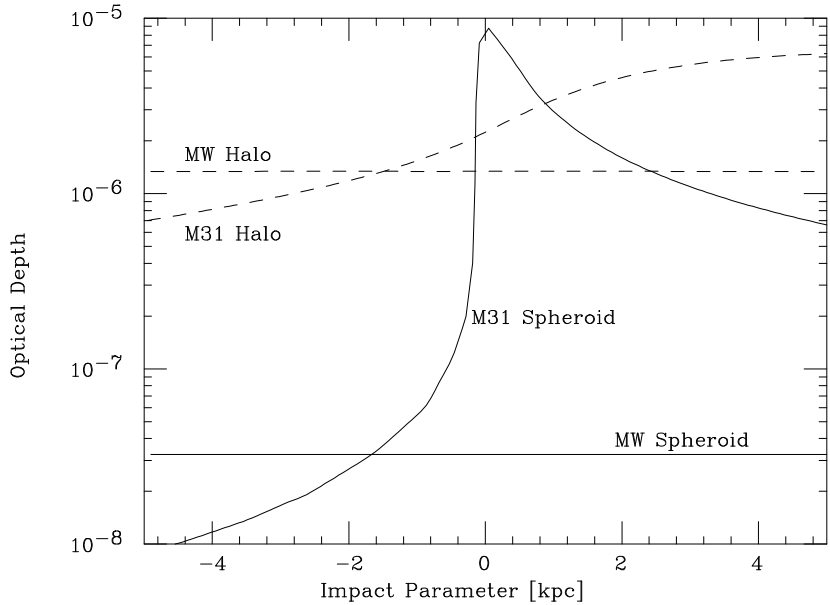,angle=90,height=10cm}}
\caption{Optical depth for ML by halo and heavy spheroid populations
of Andromeda (M31) and the Milky Way (MW), as a function of the impact
parameter between the l.o.s. to stars in the minor axis of M31 and the
M31 center.}
\label{m31.fig}
\end{figure}

Figure~\ref{m31.fig} 
  shows the optical depth dependence on $d$ for 
stars along the minor axis\footnote{Here we assume the sources to be in
the disk plane, although for small $d$ there should be corrections due
to the fact that a sizeable fraction of the sources are actually in
the M31 bulge.} and lenses with the Milky Way and Andromeda 
halo and heavy spheroid density distributions (from Ref.~\cite{gi94}). 
 If observations at 
different distances are performed, they would clearly yield valuable 
information about the distribution of the lenses.

\subsection{Infrared searches for low mass dwarfs}

There are other observations which can be helpful for identifying the 
lensing population. If the dark objects responsible for 
ML are in the form of low mass red dwarfs ($m \sim 0.1$--$0.2 
M_\odot$), which are very faint and emit mostly in the IR, or of
brown dwarfs, {\it i.e.} stars not 
massive enough to burn hydrogen but which emit anyway due to their
gravitational contraction and degenerate cooling \cite{lu86}, 
they could be detected via their 
infrared emission 
\cite{be90,ad90,da93,ne93,ke94,ba94,hu94,bo95}\footnote{The 
contribution to the diffuse X--ray background
from coronal emission of low mass dwarfs in the halo was discussed in
Ref.~\cite{ka94}.}.  The most promising brown dwarf signal 
comes from individual stars in the solar 
vicinity, while the collective emission from Galactic or 
extra--Galactic brown dwarf distributions is generally less important.

Assuming that brown dwarfs emit black body radiation, the peak flux of 
the nearest brown dwarf of a population with local density $\rho_0$ is 
\cite{ke94}
\begin{equation}
F_{max}=0.018 \left(\frac{\rho_0}{0.01 M_\odot {\rm pc}^{-3}}\right)^{2/3}
\left(\frac{m}{0.01 M_\odot}\right)^{1.04} 
\left(\frac{\tau}{10^{10} {\rm yr}}\right)^{-0.94}
\left(\frac{\kappa}{10^{-2} {\rm cm}^2 {\rm g}^{-1}}\right)^{0.225} 
{\rm Jy}.
\end{equation}
Here $m$, $\tau$, and $\kappa$ are respectively the mass, age and opacity 
of the nearest brown dwarf (${\rm Jansky}=10^{-23}$erg cm$^{-2}$s$^{-1}$Hz$^{-1}$). 
The flux peaks at the wavelength
\begin{equation}
\lambda_{max}=23 \left(\frac{m}{0.01 M_\odot}\right)^{-0.79}
\left(\frac{\tau}{10^{10} {\rm yr}}\right)^{0.31}
\left(\frac{\kappa}{10^{-2} {\rm cm}^2 {\rm g}^{-1}}\right)^{-0.075} 
\mu{\rm m}. 
\end{equation}
For a disk population of brown dwarfs, it is appropriate to take $\tau 
\simeq 5 \times 10^9$~yr and $\kappa\simeq 1$~cm$^2$~g$^{-1}$. Therefore, the present 
constraint on infrared emission from the $IRAS$ survey \cite{mo92},
$F_{max} < 0.2$~Jy for $\lambda = 12\mu$m, already sets a significant 
bound on the number  of brown dwarfs in the thin disk.

On the other hand, for the heavy 
spheroid and halo populations (for which it is 
reasonable to take $\tau =10^{10}$ yr and $\kappa=10^{-2}$~cm$^2$~g$^{-1}$, 
appropriate for older stars) the maximum fluxes are   
$\sim 0.03 (m/0.05 M_\odot)$~Jy and $0.1 (m/0.05 M_\odot)$~Jy 
respectively, therefore 
below the $IRAS$ survey sensitivity. Nevertheless, the situation will 
be improved in the near future
 by observations with the recently launched satellite 
$ISO$, and later with $SIRTF$.

The  content of low mass red dwarfs in the Galactic halo 
is also constrained by near IR searches \cite{ba94,hu94,bo95}. 
These limits again come from the effects of individual stars, as the 
diffuse emission suffers from large uncertainties from the background. 
Boughn and Uson \cite{bo95} found that no more than 3$\%$ of a
standard halo ($\rho_0^H=0.01 M_\odot$ pc$^{-3}$, $a=4$ kpc)
can be in the form of red dwarfs of $m>0.1 M_\odot$, using the result 
of a deep IR survey by Cowie et al. \cite{co90}. Pushing the local 
halo density to the minimum limit by considering a `maximum disk' 
model relaxes the limit to $6\%$. Tight bounds on the halo density in 
red dwarfs were also obtained by Hu et al. \cite{hu94} using data from 
four high latitude sky strips in the near infrared.
 Another study  by Bahcall et al. \cite{ba94} of two long exposures of a high 
latitude field taken with the Wide Field camera on the Hubble Space 
Telescope, also constrains the local 
density of red dwarfs. They   set a limit to the red dwarf
 contribution (in stars with $V-I>3$ and
$M_I>15$) to 
the local column density  of the thin disk of $\Sigma < 8 M_\odot$ 
pc$^{-2}$, and of $\Sigma < 10 M_\odot$ 
pc$^{-2}$ for stars in the thick disk. The contribution 
of main--sequence stars with $M_I>10$ to the dark halo is constrained 
to be less than 6\%. A reanalysis of this same data by Graff and
Freese \cite{gr95}, using new stellar models and parallax measurements
of low mass, low metallicity stars, tightens this last bound to be
less than 1\% of the halo density.

This means that if a significant fraction of the Galactic halo
consists  of a population of brown and red dwarfs, then their mass function 
has to cut off very sharply above $0.1 M_\odot$.

\section{Conclusions}

The idea of using ML to study the dark constituents of our Galaxy has
proved to be extremely useful. In the last few years, several
experiments started searches for ML and many surprises have resulted. 

The observations in the direction of the LMC returned rates which are
significantly smaller than those expected from a standard halo composed of
objects with masses in the range $10^{-7}$--$10^{-1}M_\odot$, but
anyway larger than those expected from the faint stars present in the
known stellar populations, so that something new has certainly been
found. If the compact objects belong to the halo, these results could 
imply that the halo has a large fraction of
heavy objects ($m>0.1M_\odot$), or a lot of gas, or is a mixture  with
non--baryonic dark matter, or alternatively that it deviates
significantly from the standard halo models due, for instance, to the
disk being close to maximal and the rotation curve actually 
falling with distance. Another plausible explanation is that a large
number of brown dwarfs and/or stellar remnants 
is present either in the Galactic spheroid or
in a thick disk, with the halo being allowed to be completely
non--baryonic.
Increased statistics will clearly allow  these facts to be put on a more
solid basis and to discriminate between the different proposed
solutions. Continuing the observations will also allow one to gain
sensitivity to longer event durations, i.e. to heavier lens masses.

Observations towards the bulge returned on the contrary more events
than  were
initially expected, implying that probably the bar in the inner Galaxy
was rediscovered using ML. Continued searches will allow one to further
constrain the geometry and mass of this Galactic component. The rates
observed may also be indicative of large amounts of material in the
disk, and this could also hint towards a smaller contribution of the
halo to the rotation curve, so that bulge results may also be relevant
for LMC searches. The large number of events observed has allowed 
identification of some particularly interesting cases of ML, such 
as events due
to binary lenses and the observation of parallax  due to the Earth's
motion. 
More events of these types, as well as proper motion
measurements in the lensing of giant stars, will allow  better
identification of the lensing agents. Also the mapping of the rates
and event
durations across the bulge are crucial for isolating the different
contributions from disk and bulge lens and source populations.

Hence, microlensing has opened a new window for Galactic astronomy,
allowing one to `see' dark objects which are otherwise
unobservable. The  first
results already had an important impact on our understanding of 
Galactic structure and of the composition of the dark halo, which still
remains one of the fundamental open problems in physics. This field is
rapidly moving, with new experiments joining the searches, so that one
can expect that a deeper, and better established, knowledge of these
fundamental issues, and possibly  some new surprises, will be obtained in
the near future.

\bigskip\bigskip

\section*{Acknowledgments}

It is a pleasure to thank  Gian F. Giudice and Alvaro de R\'ujula  for
all the work done in collaboration in this subject. We thank Andy
Gould for a careful reading of the manuscript and useful suggestions.
We also thank
D. Bennett, A. Udalski and J. Rich for allowing us to use the figures
with the results of their experiments, and  J. Kaplan,  J. Miller and 
A. Milztajn for comments.

\bigskip\bigskip

\section*{Note added} 

Just after completing this review, the first new `surprising' ML
result was announced by the MACHO collaboration\footnote{This Note is 
based on talks given by
D. Bennet, M. Pratt and W. Sutherland at the `Second International
Workshop on Gravitational Lensing Surveys', LAL, Orsay, 29--31 January
1996.}. They presented their
preliminary results from the analysis of the second year of LMC data
as well as a reanalysis of the first year data, exploiting the
experience gained with the large number of ML events observed towards
the bulge. They realized that many of the bulge events would not have
been identified with the cuts employed in the initial analysis of the
first year of LMC data, and hence used a different search strategy in the
new analysis. The results were:

\noindent {\it i)} Of the three events found initially \cite{al95e},
only one remained as a `true' ML event, the one shown in Figure
\ref{gold.fig}. Of the other two, which were low signal events, one
had  other
bumps in the light--curve during the second year, being shown then to
be just a
variable, and the other one did not pass the new cuts. However, the
star sample was enlarged by 15\% in the reanalysis of the first year
data, and with the new cuts two new events appeared during this
period.

\noindent {\it ii)} Regarding the second year, five candidate events
were found, one of which is slightly asymmetric, and hence not
completely reliable. Another of these events is due to a binary lens,
and it was possible to measure the proper motion during one of the
caustic crossings. The small value of this quantity ($\mu \simeq 2$~km/s/kpc)
suggests that the lens belongs to the LMC.  On the other hand, for the
accumulated statistics, about one event was expected due to LMC stars
acting as lenses, and the binary event appeared just in a star in the
LMC bar, while most of the remaining events are in stars outside the
bar, and hence are less likely to be due to LMC lenses.

\noindent {\it iii)} Another important fact is that the durations of
all the new events are longer (23~d to 72~d) than those found
initially  in the first year analysis, implying that the inferred lens
masses  ($m\simeq 0.1$--$1 M_\odot$) are larger than initially
suggested.

\noindent {\it iv)} A preliminary estimate of the optical depth from
the eight events is $\tau \simeq 2.5^{+1.2}_{-0.7} \times 10^{-7}$,
while removing the asymmetric event and the binary lens one (since the
lens belongs to the LMC), leads to an estimated $\tau \simeq
1.7^{+0.9}_{-0.5} \times 10^{-7}$ for the optical depth due to the
lenses which could be Galactic.

\noindent {\it v)} MACHO did a `spike' search, looking for short
duration events ($T<1$~d) leading to a couple of high amplification
points in the light--curve during one night and in the two colors. No
event was found, resulting in significant constraints for the low mass range
($m\sim 10^{-7}$--$10^{-3}M_\odot$).

\pagebreak

\appendix

\end{document}